\documentclass[12pt,preprint]{aastex}

\usepackage{emulateapj5}
\usepackage{apjfonts}

\shorttitle{A BRIDGE FROM OPTICAL TO INFRARED GALAXIES}
\shortauthors{TOTANI \& TAKEUCHI}

\submitted{Accepted for Publication in the Astrophysical Journal}

\begin{document}

\title{A Bridge from Optical to Infrared Galaxies: 
Explaining Local Properties, Predicting
Galaxy Counts and the Cosmic Background Radiation}

\author{Tomonori Totani\altaffilmark{1, 2, 3} 
and Tsutomu T. Takeuchi\altaffilmark{4, 5, 6}}

\altaffiltext{1}{
Princeton University Observatory, Princeton, NJ 08544-1001, USA \\
e-mail: ttotani@princeton.edu
}
\altaffiltext{2}{
Theory Division, National Astronomical Observatory, 
Mitaka, Tokyo 181-8588, Japan
}

\altaffiltext{3}{
Postdoctoral Fellow of the Japan Society for the Promotion of Science
(JSPS) for Research Abroad
}

\altaffiltext{4}{
Optical and Infrared Astronomy Division, National Astronomical Observatory, 
Mitaka, Tokyo 181-8588, Japan
}
\altaffiltext{5}{
Institute of Astronomy, Faculty of Science, the University of Tokyo,
Mitaka, Tokyo 181-0015, Japan
}
\altaffiltext{6}{
Research Fellow of the Japan Society for the Promotion of Science (JSPS).
}
\date{\today}

\begin{abstract}
We give an explanation for the origin of various properties observed in
local infrared galaxies, and make predictions for galaxy counts and
cosmic background radiation (CBR), by a new model extended from that
for optical/near-infrared galaxies.
Important new characteristics of this study are that (1) mass scale
dependence of dust extinction is introduced based on the 
size-luminosity relation of optical galaxies, and that (2) the big
grain dust temperature $T_{\rm dust}$
is calculated based on a physical consideration
for energy balance, rather than using the empirical relation
between $T_{\rm dust}$ and total infrared luminosity $L_{\rm IR}$
found in local galaxies, which has been employed in 
most of previous works.
Consequently, the local properties of infrared galaxies, i.e.,
optical/infrared luminosity ratios, 
$L_{\rm IR}$-$T_{\rm dust}$ correlation, and infrared luminosity
function are outputs predicted by the model, while these have been
inputs in a number of previous models. Our model indeed reproduces
these local properties reasonably well. 
Then we make predictions for faint infrared counts (in 15, 60, 90, 170, 450,
and 850 $\mu$m) and CBR by this model. 
We found considerably different results from most of previous works based 
on the empirical $L_{\rm IR}$-$T_{\rm dust}$ relation;
especially, it is shown that the dust temperature of starbursting
primordial elliptical galaxies is expected to be very high (40--80K),
as often seen in starburst galaxies or ultra luminous infrared galaxies
in the local and high-$z$ universe. 
This indicates that intense starbursts of forming elliptical galaxies
should have occurred at $z \sim 2$--3, in contrast to the previous results
that significant starbursts beyond $z \sim 1$ tend to overproduce the
far-infrared (FIR) CBR detected by {\sl COBE}/FIRAS. 
On the other hand, our model predicts that the mid-infrared (MIR) flux
from warm/nonequilibrium dust is relatively weak in such galaxies making 
FIR CBR, and this effect reconciles the {\it prima facie}
conflict between the upper limit on MIR CBR from TeV gamma-ray observations
and the {\sl COBE}\ detections of FIR CBR. The intergalactic optical
depth of TeV gamma-rays based on our model is also presented.
\end{abstract}


\section{Introduction}
{}To understand when and how stars and galaxies formed in the universe
is one of the most fundamental issues in modern astronomy.
The energy emitted from stars as a result of nuclear fusion is
radiated in two modes: direct stellar emission ranging from
optical to near-infrared (NIR) wavelengths, and emission in
mid- and far-infrared (MIR and FIR) wavelengths from
dust particles heated by stellar radiation field.\footnote{In this paper, 
the term `infrared' refers to the emission from dust particles in
the MIR and FIR including submillimeter,
but excluding direct stellar emission in the NIR.}
Faint galaxy counts and the cosmic background radiation (CBR), as well as
the properties of galaxies in the local universe, 
give us important clues to understand galaxy formation. 

In optical/NIR wavelengths, the sensitivity of existing telescopes
now reaches a depth sufficient to resolve more than 80--90\% of CBR
from galaxies (Totani et al. 2001a),
thanks to very deep optical surveys such as the Hubble Deep Fields 
(HDFs: Williams et al. 1996; Williams et al. 2000) and the NIR surveys such 
as the Subaru Deep Field (SDF: Maihara et al. 2001).
There are a number of theoretical models in various approaches
to be compared with these data, and in fact these data can be
reasonably explained in the framework of 
the Big-Bang cosmology and structure formation induced by
the cold dark matter (CDM). 
In contrast, infrared observations have not yet reached such depth 
and star formation activity hidden by dust extinction is still poorly known, 
despite the dramatic progress of observations achieved by satellites such 
as {\sl IRAS} (e.g., Soifer et al.\ 1987 and references therein) and 
{\sl ISO} (e.g.\ Puget et~al.\ 1999; Oliver et~al.\ 2000; Okuda~2000; 
for a review, see e.g., Genzel \& Cesarsky 2000). 
The SCUBA of JCMT (Holland et al.\ 1999) has opened a new window of 
submillimeter wavelengths to probe the dusty galaxies at high-$z$, 
and it seems that a considerable part of CBR from submm galaxies has been 
resolved (e.g., Smail, Ivison, \& Blain 1997; Hughes et al.\ 1998; 
Blain et~al.\ 1999; Blain et al.\ 2000; Eales et~al.\ 1999, 2000; 
and Barger et al.\ 1998, 1999). 
However, insufficient angular resolution does not allow us to identify
most of these objects in other wavelengths. 
Forthcoming projects such as {\sl SIRTF}, {\sl ASTRO-F}, SOFIA, {\sl Herschel 
Space Observatory}, {\sl NGST}, and ALMA would bring about revolutionary 
progress from this situation. 
Comparison of these data with theoretical models of infrared galaxies 
will be the key to understand the ``dark side'' of galaxy formation in 
the next decade of extragalactic astronomy.

Theoretical modeling of galaxy formation and evolution has a long history.
In optical/NIR wavelengths, most theoretical modelings can be roughly divided
into two categories: the so-called ``backwards'' approach and ``{\it ab
initio}'' approach. 
In the former approach (Tinsley 1980; Yoshii \& Takahara 1988; Fukugita et al.
1990; Rocca-Volmerange \& Guiderdoni 1990; Yoshii \& Peterson 1991, 1995; 
Pozzetti et al. 1996, 1998; Jimenez \& Kashlinsky 1999;
Totani \& Yoshii 2000), the luminosity function 
of local galaxies is used as an input to normalize the number density of 
galaxies.
The local properties of galaxies such as multi-band colors and chemical 
properties are also used to construct a reasonable model of star formation 
history and luminosity evolution of galaxies based on the stellar
population synthesis method. 
The evolution is then probed backwards into the past to predict observables 
such as galaxy counts and redshift distributions. 
The formation epoch and merging history of galaxies cannot be predicted 
in this framework, and hence they are introduced as phenomenological 
parameters that can be inferred from comparison with observational data.
In the latter approach (Kauffmann et al. 1993; Cole et al. 1994, 2000;
Somerville \& Primack 1999; Nagashima et al. 2001), on the other hand,
the formation epoch and merging history of galaxies are predicted
by the standard theory of structure formation in the CDM universe.
In these models the local properties such as luminosity
function are outputs of the model which should be compared with
observations. 
However, although the formation and evolution of dark matter halos are 
rather well understood and can be properly predicted, our knowledge about 
baryonic processes such as star formation, supernova feedback, or galaxy 
merging is still very poor, and a number of phenomenological parameters must 
be incorporated, making the comparison of a model and observed data 
rather complicated.

Both of these two approaches have been used also to make predictions
of infrared galaxy counts and infrared CBR
(see Franceschini et al. 1994 for the former, Guiderdoni et al. 1998 
and Devriendt \& Guiderdoni 2000 for the latter, and Tan, Silk, \& Balland 
1999 for a somewhat hybrid approach between the two). 
However, because of our poor knowledge and understanding of dust formation 
and emission, there are considerable difficulties in theoretically predicting 
infrared luminosity, spectral energy distribution (SED), and their evolution 
of dust emission, compared with the optical/NIR modeling based on the stellar
population synthesis method. 
This is the motivation of another ``empirical'' approach\footnote{In several 
publications this approach is also referred to a ``backwards'' approach, 
but here we discriminate between the ``backwards'' approach mentioned earlier
and the ``empirical'' approach.} in the MIR and FIR wavelengths.
In this kind of approach, the infrared luminosity evolution is 
introduced by some
phenomenological functional forms and constraints on luminosity
evolution is derived from comparison with observed data
(Beichman \& Helou 1991; Oliver, Rowan-Robinson \& Saunders 1992; 
Blain \& Longair 1993, 1996; Pearson \& Rowan-Robinson 1996; 
Malkan \& Stecker 1998, 2001; Takeuchi et al. 1999, 2001a, 2001b; 
Roche \& Eales 1999; Gispert, Lagache, \& Puget 2000; Wang \& Biermann 2000; 
Xu et al. 2001; Franceschini et al. 2001).

There is yet another approach so-called ``cosmic chemical evolution'',
in which the universe is treated as a uniform medium and the mean
gas consumption by stars in the whole universe is considered
(Pei \& Fall 1995; Fall, Charlot, \& Pei 1996; Pei, Fall, \& Hauser 1999; 
Sadat, Guiderdoni, \& Silk 2001).
The cosmic star formation history is determined by the redshift
evolution of the cosmic mass density of neutral gas and the cosmic mean
metallicity, which is inferred from quasar absorption line systems. 
Although this is a simple and beautiful approach to predict CBR
(but not galaxy counts),  several problems 
exist in this kind of approach, which could be serious especially in 
modeling infrared radiation from dust. 
First, quasar absorption systems are inevitably biased to gas-rich systems, 
i.e., systems in which star formation has been relatively inefficient, 
and hence their mean metallicity does not necessarily represent the mean 
of the whole universe. 
For example, if an elliptical galaxy has formed at $z \sim 4$, this galaxy 
has been already in passive evolution phase at $z \sim 2$.
Then it can never be observed as a quasar absorption system because of 
complete exhaustion of interstellar gas, although significant stars 
and metals have been produced.
Galaxies in dusty starburst phase may not be traced by absorption line 
systems, either, because background quasars may be completely extincted and 
cannot be observed.
Therefore, the metallicity and star formation activity
observed in absorption line systems
must be an underestimate of the true mean in the universe.
Furthermore, extinction by interstellar dust and reradiation in infrared
bands are very sensitively dependent on dust opacity and its geometrical 
distribution within a galaxy, but
this important information is completely missed.

In this paper we construct a model of infrared galaxy counts and CBR
based on the backwards approach for luminosity evolution of galaxies, 
by extending the model for optical/NIR galaxy counts (Totani \& Yoshii 2000, 
hereafter TY00) to include dust emission component in infrared bands. 
After the work by Franceschini et al. (1994, F94 hereafter), 
only a limited number of models have been published based on this approach 
covering FIR bands, while there has been a dramatic increase of quality and 
quantity of observational data in the local universe as well as for 
faint/high-$z$ galaxies. 
Although the backwards approach has a disadvantage that formation
epoch and merger history must be treated phenomenologically,
it has an advantage that the number of model parameters 
is much fewer, and connection of 
the model galaxies to the local galaxy populations is clearer, than the 
{\it ab initio} approach. 
Another reason for the relatively low activity of work in this direction is 
the difficulty of constructing realistic evolutionary models that can 
consistently describe both the optical and infrared data at the local 
universe as well as high-$z$, as argued by Franceschini et al.\ (2001). 
This is indeed what motivated more empirical studies confined to the 
infrared bands.

The aim of this paper is to explain the bulk of optical and infrared data
consistently by realistic evolutionary models of 
known and relatively normal galaxy populations at the local universe,
without introducing hypothetical populations or
parametric description of luminosity evolution. An important
improvement of our model from the F94 model is that
mass scale dependence is introduced for dust extinction and reradiation,
based on the size-luminosity relation of galaxies observed in the
optical bands. In the F94 model, there was no
physical difference in model galaxies of a given type with different
masses; massive galaxies were simply a scaled-up version of smaller 
objects. We will show that this point is essential to understand
various local properties of infrared galaxies.

Another important characteristic of this work is modeling of the
infrared SED evolution.
In most of the previous infrared models, including {\it ab initio} and 
empirical approaches, the dust SED is empirically modeled in such a way 
that the dust SED is a one-parameter family specified by 
the total infrared luminosity of galaxies. 
The relation between the dust SED and luminosity, which is observed in 
local infrared galaxies, is characterized by a gradual increase of 
characteristic dust temperature $T_{\rm dust}$ with infrared luminosity 
$L_{\rm IR}$ and is often expressed simply as a linear relation 
between $T_{\rm dust}$ and $\log L_{\rm IR}$
(e.g.\ Smith et al. 1987; Soifer \& Neugebauer 1991). 
However in this paper, we argue that this relation is not physically 
warranted for high-$z$ galaxies. In a physical sense,
the luminosity is {\it extensive},
i.e., a quantity which is dependent upon the amount
of substance present in the system. On the other hand,
the dust temperature is {\it intensive}, which is a specific 
characteristic of the system and is independent
of the amount of material concerned. Therefore these two must not be
related by only one sigle relation.
At least there should be another extensive
quantity such as the total mass of dust in a galaxy $M_{\rm dust}$. 
In the empirical $L_{\rm IR}$-$T_{\rm dust}$ relation, the extensive scale 
of an object is completely missed. 
The empirical $L_{\rm IR}$-$T_{\rm dust}$ relation of local infrared galaxies
should be considered as a projection of the plane between $L_{\rm IR}$, 
$T_{\rm dust}$, and $M_{\rm dust}$. 
Here we will construct a model on physical basis to describe the infrared 
luminosity as a two-parameter family of $M_{\rm dust}$ and $T_{\rm dust}$.
The dust temperature is physically calculated,
without using the empirical $L_{\rm IR}$-$T_{\rm dust}$ relation.
In addition to the thermal emission from big grain dust, the
warm/nonequilibrium components of dust emission from small grain
dusts and PAH features are also added to the model SED to be
consistent with the observed infrared SED of galaxies.

It is important to check that the model can successfully reproduce
the local properties of infrared galaxies, before predicting the high-$z$
quantities such as counts and CBR, since in our model all the inputs
are the optical/NIR properties of local galaxies, and those in
infrared bands are outputs predicted by the model.
We will compare the model prediction with the observed
properties of local infrared
galaxies, such as the correlation between optical and infrared
luminosity, the $L_{\rm IR}$-$T_{\rm dust}$ relation, and
infrared luminosity function.
Indeed, we will show that our model gives a reasonable explanation for 
these relations as well as the scatter along the mean relation. 
Recently Granato et al.\ (2000) have presented a very sophisticated 
semi-analytic model of infrared galaxy formation to calculate various
properties of local infrared galaxies, although predictions for high-$z$ 
galaxies have not been made yet. 
They have also found that most of the observed properties can be reproduced
in their model framework. 
However, a large number of parameters have been introduced to treat various 
physical processes, and the origin of the observed properties of local 
infrared galaxies has not been clearly discussed.
Here we try to shed light on the underlying physics of these properties
by a model with a much fewer number of adjustable parameters.

Then we will make predictions of infrared galaxy counts and CBR
by our backwards evolution model. We found that the empirical 
$L_{\rm IR}$-$T_{\rm dust}$ relation breaks down at high-$z$, especially
in the intense starbursts expected at the formation of elliptical
galaxies. The dust temperature should be much higher (40--80K) than 
predicted by mere extrapolation of the empirical $L_{\rm IR}$-$T_{\rm dust}$ 
relation.
We will discuss the implications of our results for the star formation 
history of the universe, especially the formation epoch of elliptical
galaxies. Our result also has an interesting implication for
the {\it prima facie} conflict (Stecker 2000)
between the upper limit on MIR CBR
from TeV gamma-ray observations and the FIR CBR 
detections by the {\sl COBE}\ satellite. The calculation of 
intergalactic optical depth of TeV gamma-rays based on our model
will also be presented.

The paper will be organized as follows. The model of galaxy
evolution in the optical bands, on which the infrared modeling is based,
will be summarized in \S \ref{section:model-optical}. 
Then the extension of the optical/NIR model to include the dust emission
is described in detail in \S
\ref{section:model-infrared}. Our model predicts local properties
of infrared galaxies, and they are compared with the observations
in \S \ref{section:local}. The predictions of faint infrared counts
and CBR will be presented to be compared with observed data,
in \S \ref{section:high-z}.
Implications and discussions on our results are given for several
topics in \S \ref{section:discussion}, and finally
we summarize and conclude this paper in \S \ref{section:conclusions}.
Unless otherwise stated, we will adopt a cosmological model
with $h = H_0 / (100\rm km/s/Mpc)
= 0.7$, $\Omega_0 = 0.2$, $\Omega_\Lambda=0.8$.

\section{Model of Optical/NIR Galaxies}
\label{section:model-optical}
We use the same model as that used in TY00, which is in overall agreement 
with the optical counts and photometric redshift distributions 
in HDF (TY00), and $K$-band counts, colors,
and size distributions in the SDF
(Totani et al. 2001c). 
The detail of the model is given in TY00 and here we summarize important 
properties of the model.  
Although active galactic nuclei 
(AGNs) may have some contributions to infrared galaxy counts
and CBR, most of previous studies indicate that they are not the dominant 
population in counts or CBR (e.g.\ Rigopoulou et al.\ 2000; 
Barger et al.\ 2001). 
Therefore we will consider only normal galaxies without AGNs.
Galaxies are classified into five 
morphological types of E/S0, Sab, Sbc, Scd, and Sdm, and their evolution 
of luminosity and spectral energy distributions (SEDs) is described by a 
standard galaxy evolution model in which their star formation history is 
determined to reproduce the present-day colors and chemical properties 
of galaxies (Arimoto \& Yoshii 1987; Arimoto, Yoshii, \& Takahara 1992).
These models include chemical evolution giving gas fraction $f_{\rm gas}$ 
and metallicity $Z$ as a function of time and morphological type. 
We assume that the original dust-free SED of galaxies and its evolution
do not depend on the galaxy luminosity (or mass). 
Their SED evolutions are shown in Fig.\ 1 of TY00.
The number density and morphological type mix are determined by the 
type-dependent present-day $B$-band luminosity function, 
and here we adopt that of SSRS2 
survey (Marzke et al. 1998).
Therefore, galaxies in our model are identified by morphological type and 
present-day $B$ luminosity.
All galaxies are assumed, for simplicity, to be formed at a single redshift, 
$z_F$.
We will adopt the formation redshift within a range of $z_F=2\mbox{--}5$ 
to see the sensitivity of our predictions on this parameter.

A unique characteristic of the TY00 model is that the size and luminosity 
profile of galaxies are modeled and used to calculate the detection 
completeness of high-$z$ galaxies in a realistic way based on the
formulation developed by Yoshii (1993).
Selection effects are very important in optical bands because galaxies 
are detected as extended sources and galaxy images are significantly 
affected by observational seeing and the cosmological dimming of surface 
brightness. 
In infrared bands, angular resolution is generally much larger than typical 
apparent galaxy size, and hence they are essentially observed as point 
sources. 
Therefore, extended profile of galaxies is not relevant to the detection 
probability. 
However, the size of galaxies is still very important for the opacity
of dust extinction. Here we summarize the modeling for galaxy sizes.
We use a relation between $B$-band luminosity and effective radius $r_e$ 
to calculate a characteristic size of a galaxy as a function of its
present-day $B$-band luminosity. 
The relation between the present $B$ luminosity $L_B$ and effective radius 
$r_e$ is assumed to be a power-law as $r_e \propto L_B^{2.5/p}$ for each 
galaxy type, and the normalization and $p$ are determined by fitting to 
the empirical relation observed for local galaxies. 
There is a considerable scatter in the empirical $r_e$-$L_B$ relation 
(about 0.22 in $\log r_e$), and this point will also be considered later. 
The observed $L_B$-$r_e$ relations for elliptical and spiral galaxies 
and fitting parameters to them can be seen in Fig.\ 3 of TY00.
In this paper we assume that galaxy size evolves only when number evolution 
of galaxies takes place (see below). 

A simple picture for galaxy evolution is a so-called pure luminosity
evolution (PLE) model in which there is no number evolution of galaxies
to high redshifts. 
Although some number evolution is naturally expected in the hierarchically 
clustering universe dominated by cold dark matter, it is very difficult to 
construct a realistic model of merger history of galaxies especially when 
one tries to predict infrared emission from dust.
Therefore, we use the PLE model as our baseline model. 
The difficulty of modeling realistic merger history of infrared galaxies 
is uncertainty of size evolution of galaxies by the merger processes. 
In optical/NIR bands, size of galaxies is not so important when extinction 
is not serious and the detection incompleteness is ignored. 
In infrared bands, however, emission from dust is always sensitive to dust 
optical depth that is directly correlated with size of galaxies. 
Therefore size evolution induced by merging processes should have significant 
effect on the luminosity evolution of infrared galaxies. 
However, size evolution in merging galaxies is hardly understood and very 
difficult to construct a reliable model, even in the {\it ab initio} models
where merging history of dark matter halos can be predicted rather well. 
Therefore, in this paper we will give only a simple calculation in the case 
of number evolution of galaxies, to grasp a rough behavior of the
effects of number evolution.  
We introduce a simple merger model characterized by the evolution of the 
Schechter parameters of the luminosity function, in which total 
optical luminosity density is conserved, 
i.e., $\phi^* \propto (1+z)^\eta$ and $L_B^* \propto (1+z)^{-\eta}$.
For simplicity, we assume that all types of galaxies have the same number
evolution. The size evolution is determined by assuming that 
stellar luminosity density within a galaxy
is not changed in the merger process, i.e., $L_{B} \propto r_e^3$. 
However, this is not warranted and 
should be considered as a simple toy model of 
merging galaxies.

Absorptions by interstellar dust is calculated as follows. 
The dust opacity is assumed to be proportional to the gas column density 
and metallicity, and it can be expressed by a function of restframe 
wavelength $\lambda$, redshift $z$, morphological type denoted by $k$, and 
present-day $B$ luminosity $L_{B,0}$ as
\begin{equation}
  \tau_{\rm d} (\lambda, z, k, L_{B,0}) = X(\lambda) A_{V, {\rm MW}} 
    \left( \frac{N_{\rm gas}(z, k, L_{B,0})}{N_{\rm gas, MW}} \right)
    {\left( \frac{Z(z, k)}{Z_{\rm MW}} \right)}^s \ ,
\end{equation}
where $X(\lambda) = A_\lambda/A_V$ is the extinction curve,
$N_{\rm gas}$ the gas column density, and $Z$ the metallicity of
interstellar gas.
In this paper we use the Galactic extinction curve (e.g., Pei 1992), and 
$s=1$ as a standard prescription. 
On the other hand, dependence of the extinction curve on metallicity can be
taken into account by power-law interpolations based on the solar 
neighborhood and Magellanic Clouds, with $s = 1.35$ for $\lambda
<2000${\AA} and $s=1.6$ for $\lambda > 2000${\AA} (Guiderdoni \&
Rocca-Volmerange 1987; Franceschini et al.\ 1994). We will also
try this prescription later.
Here the subscript MW denotes the value of the Milky Way at $z=0$, and we 
assume that the Milky Way is an Sbc galaxy with $L_V = 1.4 \times 10^{10} 
L_\odot$ (Binney \& Tremaine 1987).
The normalization of the optical depth is fixed by the parameter 
$A_{V, {\rm MW}}$, which will be determined later.
The column density can be expressed as:
\begin{equation}
  N_{\rm gas}(z, k, L_{B,0}) \propto 
    M_{\rm b}(k, L_{B,0}) \ f_{\rm gas}(z, k) \ r_e(z, k, L_{B,0})^{-2} \ ,
\end{equation}
where $M_{\rm b}$ is the total baryon mass of a galaxy and $f_{\rm gas}$
the gas mass fraction. 
We use the present-day $L_{B, 0}$ and mass-to-luminosity ratio 
$(M_{\rm b}/L_{B, 0})_k$ given by dust-free galaxy evolution models, 
to calculate $M_{\rm b}$. 
In a strict sense, using $(M_{\rm b}/L_{B, 0})$ 
of a dust-free evolution model is 
an approximation, because the observed $L_{B, 0}$ is already obscured by dust.
However, we use this approximation for the simplicity of numerical 
calculations. 
This approximation is reasonable provided that the optical depth of 
extinction at $B$-band in present-day galaxies is not much greater than 
unity, which is supported by observations for the majority of nearby 
galaxies. 
However, as we will argue, more massive or luminous galaxies are more dusty
and hence the above estimate of gas column density may be underestimate
for the brightest galaxies. 
This point will be discussed later.

Extinction of galactic light heavily depends on the spatial distribution
of dust in a galaxy. 
Here we adopt two extreme models of spatial distribution of dust: the 
intervening screen model and the slab (i.e., the same distribution for stars 
and dust) model. 
The attenuation factor of galactic light is 
$f_{\rm att}(\lambda, z, k, L_{B,0}) 
= \exp (-\tau_d)$ and $\{1 - \exp(-\tau_d)\}
/ \tau_d$ for the former and the latter, respectively.
Then we obtain the extincted SED of a model galaxy as 
\begin{eqnarray}
  L_\lambda(\lambda, z, k, L_{B,0}) 
    &=& f_{\rm att}^{-1}(\lambda_B, 0, k, L_{B,0}) \
    f_{\rm att}(\lambda, z, k, L_{B,0}) \nonumber \\
    &\times& L_\lambda^{\rm free}(\lambda, z, k, L_{B,0}) \ ,
\end{eqnarray}
where $L_\lambda^{\rm free}$ is the SED of dust-free model galaxies, and the 
first factor of $f_{\rm att}^{-1}$ accounts for the fact that the observed 
$L_{B,0}$ is already extincted.
The difference between the screen and slab models is not significant for
galaxies that are not so dusty (i.e., optical depth $\lesssim$ 1), but is quite
significant for dusty galaxies. 
Although the slab model seems to be more reasonable, several observations of 
starburst galaxies cannot be explained by the slab model. 
The observed correlation between power-law index of UV spectra and Balmer 
line ratio of starburst galaxies indicates that the observed reddening of 
starburst galaxies is larger than expected from the slab model, and at least 
some fraction of dust should behave like a screen (Calzetti, Kinney, \& 
Storchi-Bergmann 1994, see also Gordon, Calzetti, \& Witt 1997). 
Although the screen model may seem unreasonable, it is rather reasonable 
that a fraction of dust behaves like a screen with a distribution more
extended than stars, since strong wind from starbursts could blow
out the dust particles along with interstellar gas. 
This could occur in the scale of star forming regions within a galaxy and/or
the whole scale of a starbursting galaxy. Some ultra luminous infrared
galaxies (ULIRGs) are known to
have infrared luminosity which is more than 100 times higher than
optical/NIR luminosity. In the slab model, this requires an enormous
amount of dust with $\tau_{\rm d} \gtrsim 100$, while in the 
screen model dust amount can be more modest.
Therefore the screen model may be better than the slab model as a 
phenomenological prescription.
We then use the screen model as a standard, and also use the slab model 
to see the dependence on the dust distributions. 

The model described here
has already been compared comprehensively with the counts and 
redshift distributions of galaxies observed in optical and NIR bands 
(Totani \& Yoshii 2000; Totani et al. 2001c). 
It has been found that all data can be reasonably explained by this model 
assuming a $\Lambda$-dominated flat cosmology. 
The NIR data can be explained well by the PLE model
(Totani et al. 2001c), while the optical data suggest some modest
number evolution ($\eta \sim 1$) (Totani \& Yoshii 2000). The
apparently discrepant results between optical and NIR bands
may reflect dependence of number evolution on galaxy types, i.e.,
almost no number evolution for elliptical galaxies while a stronger
number evolution for late-type galaxies
(see Totani et al. 2001c for more detailed discussion).

\section{Extension to Infrared Galaxies}
\label{section:model-infrared}
\subsection{Determination of Effective Dust Temperature}
In the optical modeling described above, we have already calculated the 
absorption of stellar light by dust in a galaxy of given type at any 
redshift (or time). 
Therefore we can straightforwardly obtain the total infrared luminosity 
emitted by dust particles as
\begin{eqnarray}
  L_{\rm IR} (z, k, L_{B,0}) &=& \int d\lambda  \
    f_{\rm att}^{-1}(\lambda_B, 0, k, L_{B,0}) \nonumber \\
    &\times& [1 - f_{\rm att}(\lambda, z, k, L_{B,0})]  \
    L_\lambda^{\rm free}(\lambda, z, k, L_{B,0}) \ .
\end{eqnarray}
Then the task remained to predict infrared flux at a given observing band 
is to model the SED of dust emission. 
In a number of previous papers, an empirical relation between the infrared 
SED (or characteristic dust temperature $T_{\rm dust}$) and $L_{\rm IR}$ has 
been utilized here. However, as mentioned in Introduction,
it is physically unreasonable to assume that the luminosity is 
determined by only one quantity, i.e., temperature. 
The total luminosity is an extensive quantity that is proportional to dust 
mass, provided that galaxies are transparent for dust emission.
If a galaxy is simply scaled up in mass, the infrared luminosity should also
increase but the temperature should not, 
which is an intensive quantity determined
by the local physical state of dust particles.
We have no reasonable physical explanation for this empirical relation
so far, and 
it is not warranted at all that this can be applied to high-redshift galaxies. 

Rather, dust temperature should be determined by an energy balance condition; 
i.e., temperature should be a value at which dust emissivity is consistent 
with total amount of stellar light absorbed by dust. 
We assume that total infrared luminosity is dominated by thermal radiation
of big grain dust, which is well described by the so-called graybody 
spectrum, i.e., blackbody spectrum multiplied by the dust emissivity that is 
described by a power of frequency as $\epsilon \propto \nu^\gamma$, 
in the FIR regime. 
This is a good approximation for local galaxies, and we will find that it 
holds also for high-$z$ galaxies whose dust temperature is higher.
Then the energy balance condition can be written as:
\begin{equation}
  L_{\rm IR} = L_{\rm IR, *} f_{\rm d}
    \left( \frac{M_{\rm dust}}{M_{\rm dust, *}}
    \right) \left( \frac{T_{\rm dust}}{T_{\rm dust, *}} \right)^{4+\gamma} \ ,
\label{eq:energy-balance}
\end{equation}
where $M_{\rm dust}$ is the total dust mass in a galaxy and 
the dust emissivity index $\gamma$ is typically 
$\gamma$ = 1--2 (Boulanger et al.\ 1996; Hirao et al.\ 1996; 
Calzetti et al.\ 2000; Dunne et al.\ 2000). 
(The case of $\gamma = 0$ corresponds to the blackbody spectrum.) 
Throughout this paper we use $\gamma$ = 1.5.
The factor $f_{\rm d}$ is a mass fraction of dust heated by stellar light.
In the case of the slab dust, all dust should be heated equally,
while in the case of the screen dust only a part of dust can be heated
by penetrating light. Therefore,
\begin{eqnarray}
f_{\rm d} = \left\{  
\begin{array}{cl}
  1, & {\rm (slab)}  \\
  \displaystyle{ 
  \frac{1 - \exp [ -\tau_{\rm d}(\lambda_{\rm eff})]}{\tau_{\rm d}
  (\lambda_{\rm eff})} }, & {\rm (screen)}
\end{array}  \right.
\label{eq:f_d}
\end{eqnarray}
where the characteristic wavelength 
$\lambda_{\rm eff}$ of absorbed stellar light is defined as:
\begin{equation}
  \lambda_{\rm eff} = \frac{\int d\lambda \ \lambda \
    [1 - f_{\rm att}(\lambda)] L_\lambda^{\rm free}(\lambda) }{\int d\lambda \
    [1 - f_{\rm att}(\lambda)] L_\lambda^{\rm free}(\lambda)} \ . 
\end{equation}
The subscript * denotes the reference point to fix the normalization of  
eq. (\ref{eq:energy-balance}). 
We take this reference point as $L_{\rm IR, *} = 10^{10} h^{-2} L_\odot$ and
$T_{\rm dust, *} = 19$ K for a galaxy type Sbc at $z=0$, to be consistent with 
the observed $L_{\rm IR}$ and $T_{\rm dust}$ relation.
The relative dust mass ratio $(M_{\rm dust}/M_{\rm dust, *})$ can be 
calculated for any model galaxy assuming that dust-to-metal ratio is constant, 
i.e., $M_{\rm dust} \propto f_{\rm gas} M_{\rm b} Z$. 
In this way we obtain $T_{\rm dust}$ for any model galaxy.

\subsection{The Infrared SED Templates}
\label{section:SED}
The overall shape of a galaxy SED is basically determined by dust size 
spectrum and strength of radiation field.
Here we should note that the entire range of infrared SED of dust emission 
cannot be described by a simple graybody spectrum with a single temperature. 
While FIR peak of the infrared SED is well described by a single-temperature 
graybody spectrum of cold, big grain dust, the MIR emission is dominated by 
radiation from heated small grains and PAHs that are not in thermal 
equilibrium with ambient radiation field. 
Because of their small heat capacity compared with the heating photon
energy, these dust grains are heated 
stochastically by one or two photons, and their `temperature' temporally 
variates violently (e.g., Purcell 1976; Draine \& Anderson 1985; 
Siebenmorgen, Kr\"{u}gel, \& Mathis 1992; Draine \& Li 2001; 
Li \& Draine 2001).
This phenomenon makes the MIR continuum very broad, and causes a power-law 
like spectrum.
Thus, it may be improper to describe the MIR continuum as a superposition of 
multi-temperature blackbody spectra.

Though the MIR continuum is dominated by the contribution from non-equilibrium
dust grains, it is well correlated with temperature of big grain dust, 
$T_{\rm dust}$ (Shibai, Okumura, \& Onaka 2000; Dale et al. 2001).
It enables us to construct the overall MIR--FIR SEDs as a unique
function of $T_{\rm dust}$.
We utilize the observed correlations between MIR band fluxes (Dale et 
al.\ 2001) and construct the MIR regime of the model SED in an empirical way.
This is a standard prescription used for the dust SED modeling in a number of 
previous papers. 
This is equivalent to assuming that the local physical state of dust 
particles is specified by one parameter, i.e., characteristic temperature.
It is physically not unreasonable, though not warranted, to expect that this 
holds also for high-$z$ galaxies, unless the composition 
or size distribution of 
dust particles in high-$z$ galaxies is drastically different from local 
galaxies. 
The point different from previous models is that we do not simply relate 
$T_{\rm dust}$ to the infrared luminosity $L_{\rm IR}$, but derive 
$T_{\rm dust}$ by the energy balance condition. 

Details of the construction of the MIR SED are as follows. 
The spectrum of thermal, big grain dust particles is described by
a graybody spectrum with $T_{\rm dust}$. 
The MIR continuum coming from small particles is described by a broken 
power law with two different indices, which are determined by the observed 
flux at 12 and 25 $\mu$m compared with that in $60\;\mu$m at which the 
graybody is dominant. 
These relations are:
\begin{eqnarray}
  \log \frac{S_{25}}{S_{60}} = -0.026 \,T_{\rm dust} + 0.367 \;,
\end{eqnarray}
and 
\begin{eqnarray}
  \log \frac{S_{12}}{S_{25}} = -1.75 \log \frac{S_{60}}{S_{100}} - 0.866 \;,
\end{eqnarray}
where $S_{\rm x}$ is the flux density of infrared galaxies in units of 
$\rm Wm^{-2}Hz^{-1}$ at $\lambda$ = x $~\mu$m, and the temperature
$T_{\rm dust}$ is in unit of K.
In addition to this, the PAH features are taken from spectroscopic 
data obtained by {\sl ISO}\ SWS (Dale et al. 2001), and their strength is
determined by the flux at 6.75~$\mu$m compared with the continuum
emission as:
\begin{eqnarray}
  \log \frac{S_{6.75}}{S_{15}} = \left\{
    \begin{array}{ll}
      -1.57 \log{S_{60}}/{S_{100}} - 0.618 & ({S_{60}}/{S_{100}}>0.4)\,,\\
      -0.104 & ({S_{60}}/{S_{100}} \leq 0.4) \ .
    \end{array}
  \right.
\end{eqnarray}
Thus we obtain the MIR--FIR SEDs of galaxies with various dust temperature.
Here we summarize the shape and frequency range of each component as follows: 
i) a power-law component ($1.0 \times 10^{12} \mbox{--} 1.0 \times 10^{13}$ 
[Hz]). 
The longer wavelength edge of this component is set to decline as a power law 
$\propto \nu^{2 + \gamma}$, and shorter wavelength edge is set to decline 
exponentially.
ii) a PAH continuum obeying a power law ($1.0 \times 10^{13} \mbox{--} 
3.0 \times 10^{13}$ [Hz]).
This component is connected to the above one smoothly.
iii) a PAH band emission component ($2.5 \times 10^{13} \mbox{--} 1.0 
\times 10^{14}$ [Hz]).
Finally we added radio continuum to the IR SED along with the tight radio-FIR 
correlation (e.g.\ Helou, Soifer, \& Rowan-Robinson 1985; 
Condon 1992),
\begin{eqnarray}
  \log S_{\rm 1.4 GHz} = \log \left(0.867 S_{60}+0.336 S_{100}\right) - 2.3 \;.
\end{eqnarray}

\begin{figure}
\plotone{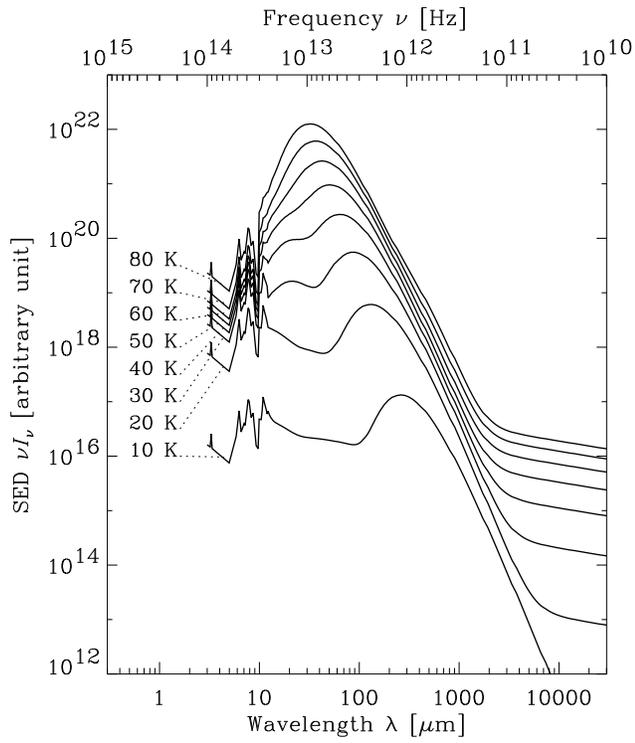}
\caption{The infrared spectral energy distribution (SED) of model
galaxies, for various values of the temperature
of big grain dust in thermal equilibrium (indicated in the figure).
The nonthermal emission coming from small grain
dust and PAH features are also taken into account. 
}
\label{fig:SED-schem}
\end{figure}

Figure~\ref{fig:SED-schem} shows our model SED for various values of 
$T_{\rm dust}$. 
A trend is seen that the MIR component coming from nonequilibrium dust 
emission becomes weaker compared with the graybody emission from big grain 
dust, with increasing temperature. 
This trend is reasonable because the high temperature is equivalent
to stronger flux of heating radiation field. 
When the heating intensity is stronger, the heating photons are more 
frequently absorbed by small dust grains, and hence more dust particles
become in thermal equilibrium, making the spectrum closer to the graybody.  
There is also a large body of evidence that the PAH are likely to be 
destroyed, or that their emission is severely damped, in regions of high 
heating intensity (Dale et al. 2001 and references therein). 
These trends have an important implication for the 
compatibility of MIR CBR and 
TeV gamma-ray observations, which will be discussed in \S \ref{section:TeV}.

\begin{figure}
\plotone{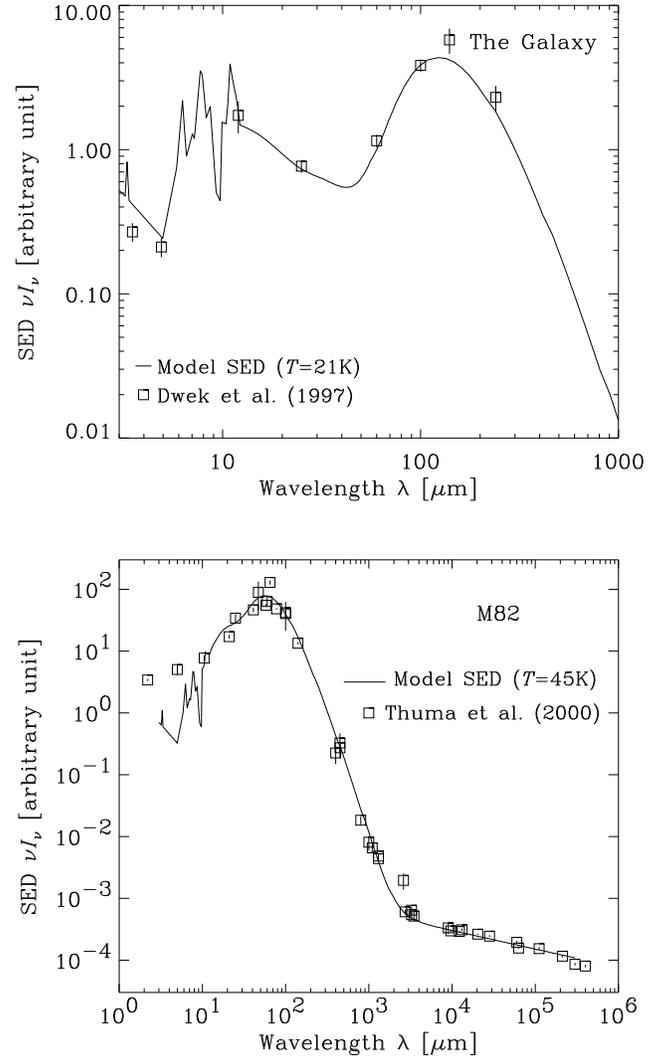}
\caption{Infrared spectral energy distribution (SED) of the Milky Way
(top panel) and M82 (bottom panel). The model curves are fit by our
model SED described by a characteristic temperature of cold, big
grain dust (indicated in the panels), 
while the nonthermal emission coming from small grain
dust and PAH features are also added. 
The observed data are from
Dwek et al.\ (1997) for the Milky Way, and Thuma et al.\ (2000) for M82.
}
\label{fig:SED1}
\end{figure}

\begin{figure}
\plotone{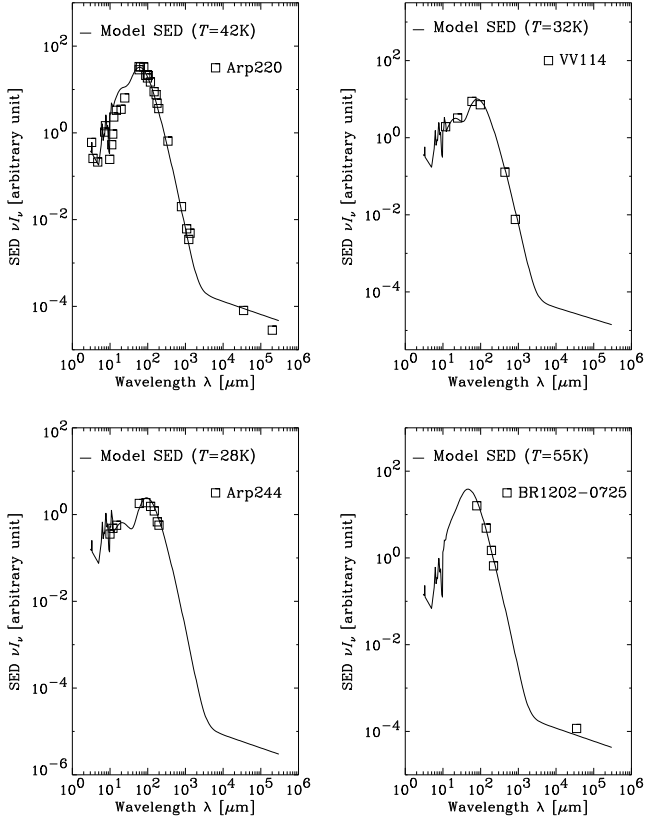}
\caption{The same as Fig. \ref{fig:SED1}, but for other four 
nearby infrared galaxies. 
The observed data are taken from Klaas et al.\ (1997) for Arp 220 and Arp 244, 
Frayer et al.\ (1999) for VV~114, and Yun et al.\ (2000) for BR~1202$-0725$,
respectively.
}
\label{fig:SED2}
\end{figure}

Figure \ref{fig:SED1} shows the model infrared SEDs for $T_{\rm dust} = 21$ 
and 45~K, corresponding to the cases of the Milky Way and M82. 
Our model is in good agreement with the observed SEDs in both cases. 
Figure~\ref{fig:SED2} is the same, but showing the fit of our model to 
other four infrared galaxies with various values of $T_{\rm dust}$. 
These results indicate that our model of infrared SEDs is in
reasonable agreement 
with observed infrared galaxies in a wide range of dust temperature.

Now we can predict various observable quantities of infrared galaxies, i.e., 
local properties 
as well as faint galaxy counts and CBR composed by
high-redshift galaxies, based on the model of optical galaxies described in 
the previous section. 
The only one parameter not yet determined is $A_{V, {\rm MW}}$, the overall 
normalization of dust extinction.
Increasing $A_{V, {\rm MW}}$ makes galaxies more dusty and hence more
luminous at infrared wavebands. 
Infrared galaxy counts at a fixed flux then increase with $A_{V, {\rm MW}}$. 
In this paper we set this parameter to a value at which bright infrared
galaxy counts at 15, 60, and 90 $\mu$m are consistent with the {\sl IRAS}\ 
counts (see \S \ref{section:counts}). 
This is found to be $A_{V, {\rm MW}} \sim 0.7$ for our baseline model. 
However, it is important to check whether the model is consistent with
observations for local infrared galaxies, before we discuss the result of 
counts and the CBR. This will be presented in the next section.

\section{Local Properties in Infrared: Model versus Observations}
\label{section:local}

Here we compare our model predictions with the properties
of local galaxies observed by {\sl IRAS}. 
We use the sample of the PSC$z$ survey of 15411 {\sl IRAS} galaxies selected 
by 60~$\mu$m flux (Saunders et al.\ 2000). 
Out of this sample, we further selected 12324 galaxies detected in the 
100~$\mu$m band as well, for which the bolometric infrared
luminosity can be reliably estimated. 
Systematic and reliable optical flux measurement is not available for 
all of these galaxies (see the ReadMe file of the PSC$z$ catalog of 
Saunders et al. 2000 for detail). 
We use the Zwicky/RC3 magnitude as $B$-band flux, which is
available for 5803 optically bright galaxies.

\subsection{Optical/Infrared Luminosity Ratio}
\begin{figure}
\plotone{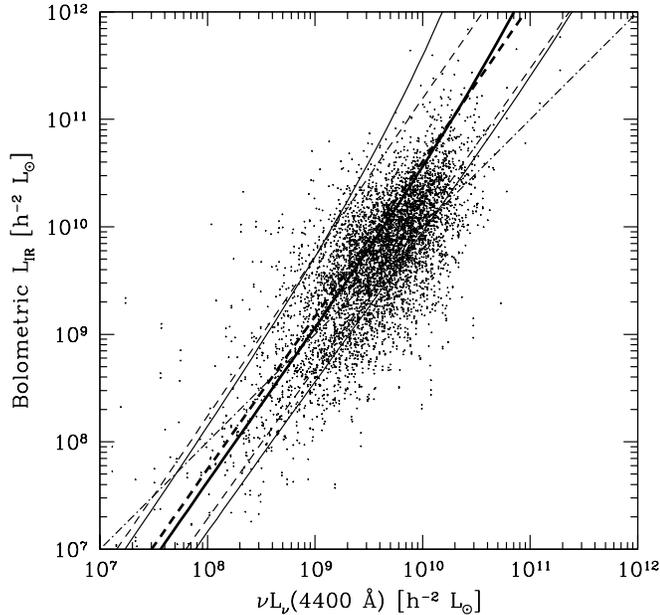}
\caption{Correlation plot of the bolometric infrared luminosity
versus optical luminosity ($B$-band) of galaxies. 
The data are from Saunders et al.\ (2000). 
The dot-dashed line is that corresponds to $L_{\rm IR} /
\nu L_\nu (B) = 1$. The thick solid and dashed lines are the
prediction of our model assuming the screen and slab distribution of
dust, respectively. The thin solid and dashed lines are the same 
with the thick lines, but in the cases where the size of galaxies
are shifted by $\pm 1 \sigma$ of observed dispersion in $\log r_e$.
}
\label{fig:L-IR-B}
\end{figure}

Figure~\ref{fig:L-IR-B} shows an $L_B$ ($\nu F_\nu$ at 4400{\AA}) versus 
$L_{\rm IR}$ (bolometric) plot. 
The data points are 
the 5803 galaxies with available $B$-band magnitudes. 
A clear trend can be seen that more luminous galaxies are more dusty, i.e., 
showing high $L_{\rm IR}/L_B$ ratio.
On the other hand, the thick solid and dashed lines are our prediction 
averaged over all the four spiral galaxy types at $z=0$, for the screen and 
slab dust, respectively.  
Here galaxy types are averaged by their relative proportions calculated from 
the type-dependent $B$-band luminosity function. 
It should be noted that elliptical galaxies are not included in this 
prediction because they do not have dust and hence infrared emission at 
$z=0$ in our model.
The trend of increasing $(L_{\rm IR}/L_B)$ ratio with luminosity is well 
reproduced by the model. 
This trend stems from the adopted $B$ luminosity versus size relation; dust 
opacity, which is proportional to the metal column density, increases with 
$L_B$ when the locally observed $L_B$-$r_e$ relation is taken into account. 
Consequently, more massive galaxies emit their radiation more in infrared 
bands. 
There is a considerable scatter in the $L_B$-$r_e$ relation, and this should
produce scatter in the optical depth and hence a scatter along the mean 
$L_{\rm IR}$-$L_B$ relation.
The thin-solid and thin-dashed curves are 
the predictions when the $L_B$-$r_e$ relation is shifted by 
$\pm 1\sigma = 0.223$ in $\log r_e$ from the mean relation,
where the line markings are the same with those of thick lines.
The scatter is also in reasonable agreement with the observed scatter.

There is some offset between the prediction and mean observed
$L_{\rm IR}$-$L_B$ relation. 
We suggest that it is due to the incompleteness of the sample; the RC3
$B$-magnitude is available only for galaxies bright in optical bands,
which are about half of the whole infrared sample of 12324 galaxies.
The galaxies without $B$-magnitudes are then 
expected to be less luminous in optical bands. 
This effect could lead to the lack of infrared luminous galaxies compared with 
the model prediction in Fig.~\ref{fig:L-IR-B}.

\subsection{Dust Temperature versus Infrared Luminosity}
\begin{figure}
\plotone{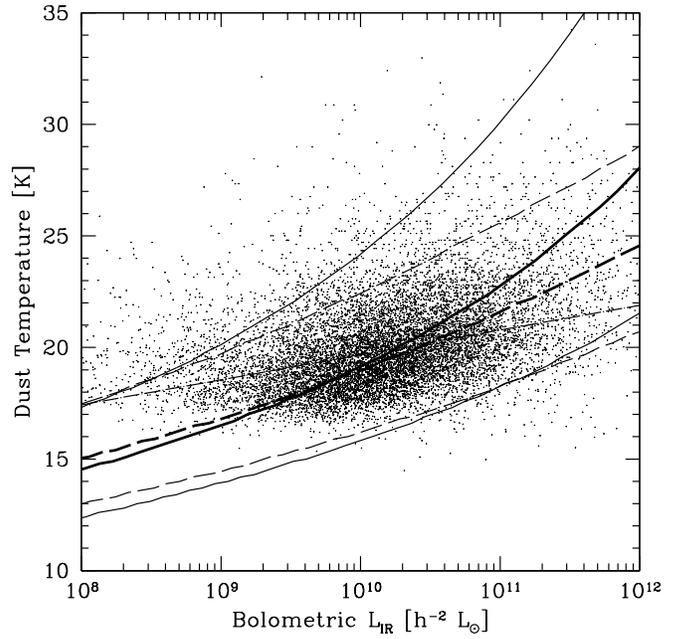}
\caption{Correlation plot of the bolometric infrared luminosity
versus temperature of cold, big grain dust estimated from
60 and 100 $\mu$m fluxes. 
The data are taken from Saunders et al.\ (2000).
The thick solid and dashed lines are the
prediction of our model assuming the screen and slab distribution of
dust, respectively. The thin solid and dashed lines are the same 
with the thick lines, but in the cases where the size of galaxies
are shifted by $\pm 1 \sigma$ of observed dispersion in $\log r_e$.
The dot-dashed line is a simple linear fit to the data, which is
used to determine the dust temperature based on 
the empirical $L_{\rm IR}$-$T_{\rm dust}$ relation (eq. 
\ref{eq:empirical-L-T}).
}
\label{fig:L-IR-T}
\end{figure}

Figure~\ref{fig:L-IR-T} shows the infrared bolometric luminosity
versus dust temperature plot. 
The data points are the 12324 galaxies in the {\sl IRAS}~PSC$z$ sample with 
100-$\mu$m flux available. 
Dust temperatures of the {\sl IRAS} galaxies are estimated from their 
$S_{60}/S_{100}$ flux ratio based on the temperature measurement of the
Galactic dust by {\sl IRAS} and {\sl COBE} (Nagata et al.\ 2001, in 
preparation).
The thick/thin solid and dashed lines are the model predictions, with the 
same line markings corresponding to model predictions shown in 
Fig.~\ref{fig:L-IR-B}. 
A trend of gradual increase of mean $T_{\rm dust}$ with $L_{\rm IR}$,
as well as the scatter along the mean relation, is again well reproduced by 
the present model. 
This also stems from the fact that more luminous galaxies are more dusty, 
as argued in the previous subsection. 
This fact requires that $L_{\rm IR}$ should increase faster than a simple 
scaling to $M_{\rm dust}$, when the mass of galaxies is increased.
Considering the energy balance of $L_{\rm IR}
\propto M_{\rm dust}T_{\rm dust}^{4 + \gamma}$, the dust
temperature must gradually
increase with $M_{\rm dust}$ or $L_{\rm IR}$.

There is a clear lack of low temperature galaxies with $T_{\rm dust}
\lesssim 16$K in the {\sl IRAS}~sample. 
We suggest that this sudden lack may come from selection effects. 
As we mentioned above, the dust temperature is calculated based on the 
60-$\mu$m and 100-$\mu$m fluxes.
However, the sample is selected at {\sl IRAS}~60-$\mu$m bandpass, and very 
cold galaxies have an IR SED peak at wavelength much longer than 60-$\mu$m.
In addition, such galaxies have low luminosity in general.
Therefore, cold galaxies are easily dropped by the sampling procedure in 
the sample used here.
This is likely to be the cause of the sudden cutoff of the data on the diagram.

\subsection{Infrared Luminosity Function}
\label{section:LF}
\begin{figure}
\plotone{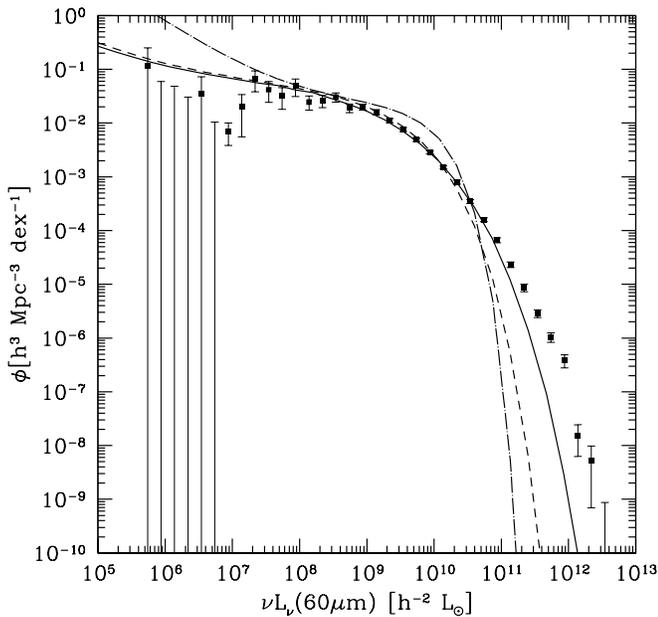}
\caption{Infrared luminosity function of galaxies at 60$\mu$m.
The data are taken from Saunders et al. (1990).
The solid and dashed lines are the
prediction of our model assuming the screen and slab distribution of
dust, respectively. Considering the uncertainty in the normalization 
of luminosity function, the normalization of model curves are
set to match the data in the range of $\nu L_\nu(60\mu{\rm m}) \leq
10^{11} h^{-2} L_\odot$. [More luminous galaxies are removed from
normalization fit to avoid contamination of AGNs which is expected
to be significant for ultraluminous galaxies 
(Veilleux, Kim, \& Sanders 1999).]
The dot-dashed line is the prediction from
the Schechter-type $B$-band luminosity function when
the infrared luminosity is simply proportional to the $B$-band luminosity
in all galaxies as $\nu L_\nu(B)/\nu L_\nu(60\mu{\rm m})$=1.
}
\label{fig:IR-LF}
\end{figure}

Finally we compare the model prediction of local infrared luminosity
function (at 60$\mu$m) to the estimated one from observations 
(Saunders et al. 1990), in Fig. \ref{fig:IR-LF}.
It is well known that the infrared luminosity function is considerably
different from that of optical galaxies that is well fitted by the Schechter 
function. The dot-dashed line is the Schechter-type luminosity function 
translated from the $B$-band luminosity function assuming 
$\nu L_\nu(B)/\nu L_\nu(60\mu{\rm m})=1$.
In sharp contrast to the rapid exponential decline of the optical luminosity 
function above $L^*$, infrared luminosity function extends well beyond $L_{60}
\sim 10^{10} L_\odot$ where $L_{60}$ is $\nu L_\nu$ at 60$\mu$m
(Fig.~\ref{fig:IR-LF}). 
Despite this intrinsic difference between the two, our model prediction 
with the screen dust, which is based on the $B$-band luminosity function 
having the Schechter form, well reproduces the shape of 60$\mu$m luminosity 
function up to $L_{60} \sim 10^{11} L_\odot$. 
Here, the infrared luminosity function is calculated as
\begin{equation}
\phi_{60} (L_{60}) = \sum_k \left( \frac{dM_B}{d\log_{10}L_{60}} \right)_k
\phi_{B, k} (M_B) \ ,
\end{equation}
where $\phi_{B, k}$
is the $B$-band luminosity function of the $k$-th type galaxies.
This is again due to the fact that more luminous galaxies are more dusty; 
when optical luminosity is increased, infrared luminosity increases more 
rapidly than that, and hence an optical luminosity range is significantly 
broadened when it is translated into an infrared luminosity range. 
This effect is further strengthened for the infrared luminosity function at 
$60 \mu$m, since the gradual increase of temperature with bolometric 
luminosity makes the increase of 60$\mu$m luminosity even faster.

The models fail to reproduce the data at ultra-luminous region of 
$L \gtrsim 10^{12} L_\odot$, possibly because of the contribution of 
peculiar starbursting galaxies under galaxy interactions, or AGNs. 
In fact observations suggest that the fraction of AGNs increases with infrared 
luminosities (Veilleux, Kim, \& Sanders 1999). Our model does not
take into account these rare populations of ULIRGs
at the local universe, since their contribution to
counts or CBR is expected to be small (Rigopoulou et al. 2000;
Barger et al. 2001). Instead, we try to explain bulk of the infrared
data by the population of normal galaxies at the local universe
and their ancestors at high redshifts.
It should also be noted that the approximation about mass-to-light ratio 
$(M_{\rm b}/L_B)$ 
breaks down for very bright and dusty galaxies, as mentioned in 
\S \ref{section:model-optical}. 
Therefore our calculation is quantitatively unreliable and
could be an underestimate at the brightest 
part of the infrared luminosity function. 

Recently the dust mass function of local galaxies is derived by
Dunne et al. (2000), based on infrared-submillimeter SEDs.
They found that the dust mass function can be fitted by the
Schechter function, in contrast to the 60$\mu$m luminosity
function. This is also consistent with the expectation of our model;
the dust mass is roughly proportional to the $B$-band luminosity
unless the optical light is heavily extincted, and hence the dust
mass function should be the same shape with the optical luminosity 
function.

\section{High Redshift Infrared Galaxies}
\label{section:high-z}
\subsection{Evolution of Infrared Luminosity and SED}
\begin{figure}
\plotone{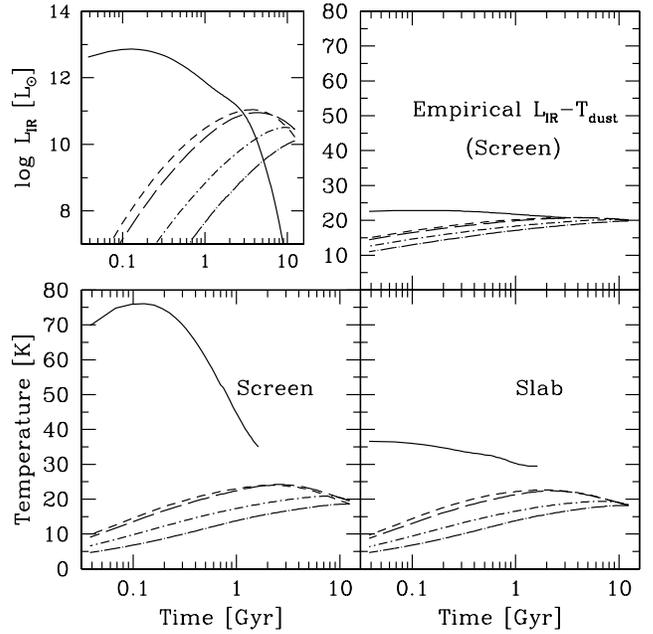}
\caption{The evolution of infrared luminosity and characteristic
temperature of big grain dust predicted by our model as
a function of time after the formation,
for five types of galaxies: E/S0 (solid), Sab (short-dashed),
Sbc (long-dashed), Scd (dot-short-dashed), and Sdm (dot-long-dashed).
The upper-left panel shows the evolution of total infrared luminosity
of our model assuming the screen-type dust. The upper-right panel
shows the temperature evolution predicted when one uses the
empirical $L_{\rm IR}$-$T_{\rm dust}$ relation observed for local
infrared galaxies. The lower-left and -right panels
show the predictions of dust temperature evolution based on our
model taking into account the energy balance between absorption
and re-emission of stellar light, for the screen- and slab-type
dust distributions, respectively. All models are for galaxies
whose luminosity is $M_B = -20$ at $z=0$.
}
\label{fig:LT-ev}
\end{figure}

In the following part of this paper we will make predictions for high-$z$
infrared galaxies such as faint galaxy counts and CBR.
First we show the evolution of infrared luminosity and
big-grain dust temperature ($T_{\rm dust}$)
of model galaxies as a function of
time after the formation in Fig.\ \ref{fig:LT-ev}. These models are
constructed to reproduce the observed optical SEDs and chemical
properties at the typical age of present-day galaxies ($\gtrsim$10Gyr).
For $T_{\rm dust}$ the three cases of dust temperature modeling are shown in
the three panels: the screen
dust, slab dust, and using the empirical relation between
$L_{\rm IR}$ and $T_{\rm dust}$ of the local galaxies. 
Here, we used the best-fit relation to the observed data as:
\begin{equation}
T_{\rm dust} = 1.1 \log_{10}( L_{\rm IR} / L_\odot ) + 8.5 \ \rm K
\ ,
\label{eq:empirical-L-T}
\end{equation}
for the empirical $L_{\rm IR}$-$T_{\rm dust}$ relation
(shown as a dotted line in Fig. \ref{fig:L-IR-T}).
It should be noted that our evolution model depends on the mass scale of
galaxies. Here we are showing the models for galaxies whose 
present-day $B$-band luminosity is $M_B = -20$, as typical giant
galaxies. It can 
be seen that dust temperature evolution is much stronger than
expected from the empirical $L_{\rm IR}$-$T_{\rm dust}$ relation,
especially for primordial elliptical galaxies. The peak dust temperature
during the starburst phase of elliptical galaxies in the slab-type dust
model is $\sim$30--40K, and even higher temperature of $\sim$40--80K
is found for the screen dust model, while the 
the empirical $L_{\rm IR}$-$T_{\rm dust}$ relation predicts 
$T_{\rm dust} \lesssim$ 25K in all time. This 
must have significant effect on the infrared SED of primordial 
elliptical galaxies, and hence on faint counts and CBR.

In the primordial elliptical galaxies, the dust opacity is so high
and hence $\tau_{\rm d} \gg 1$ for most of optical light. Therefore,
the total absorbed light, i.e., $L_{\rm IR}$ is not much different
between the slab and screen models. Rather, the difference of the two
mainly comes from the difference of amount of dust irradiated by
stellar radiation. In the case of sceen dust model with $\tau_{\rm d} 
\gg 1$, only
$f_{\rm d} \sim \tau_{\rm d}(\lambda_{\rm eff})^{-1}$ of total amount of
dust can be heated, while all dust is heated in the slab model
(eq. \ref{eq:f_d}).
Consequently the temperature of the screen dust model must be
higher to achieve higher luminosity per unit dust mass.
It should be noted that the temperature estimate in the screen model
is quite robust, in a sense that it does not depend on 
the total amount of dust. Once $\tau_{\rm d} (\lambda_{\rm eff})
\gg 1$ is achieved,
the dust-mass dependence is canceled between $f_{\rm d}$ and
$M_{\rm dust}$ in the energy balance equation
(eq. \ref{eq:energy-balance}). Later, we will find
that the screen model shows much better agreement with counts and CBR,
than the slab model.

The high temperature which we obtained here for primordial elliptical
galaxies should be compared with some observational estimates
of dust temperature for high-$z$ as well as local starburst galaxies.
It should be noted that the dust temperature obtained from a fit
to observed spectra depends on several other parameters, such as
the emissivity index $\gamma$ and opacity of a galaxy to FIR
radiation. A same observed SED may be fitted with a higher $T_{\rm dust}$
when a smaller value of $\gamma$ is assumed or the opacity to infrared
radiation ($\tau_{\rm IR}$)
is significant. On the other hand, the model temperature
derived here is an effective value derived from the total energy
balance between dust mass and bolometric luminosity assuming $\gamma = 1.5$
and low opacity to infrared light. Therefore,
our result should be compared with observational estimates of
temperature derived under the same assumptions of $\gamma \sim 1.5$ 
and low opacity to FIR light.
Keeping this point in mind, known high-$z$ luminous starburst galaxies
seem to have a mean temperature of about 50K, under the above
conditions for $\gamma$ and infrared opacity, and some of them
show a high temperature of $\sim 70$K (Ivison et al. 1998a, b;
Benford et al. 1999). Recently a comprehensive study for the SED
of local ULIRGs is published (Klaas et al. 2001), and it again 
shows that the ULIRG dust temperature ranges from $\sim$40 to 
$\sim$80K, with $\gamma \sim 1.6$ and $\tau_{\rm IR} (100\mu{\rm m}) \sim$
0.5--5. These observations then indicate that the temperature
derived by our model for primordial elliptical galaxies is
in a reasonable range of typical known starburst galaxies, although it
is much higher than the simple extrapolation of the mean
$L_{\rm IR}$-$T_{\rm dust}$ relation of local galaxies.

\subsection{Infrared Galaxy Counts}
\label{section:counts}
\begin{figure}
\plotone{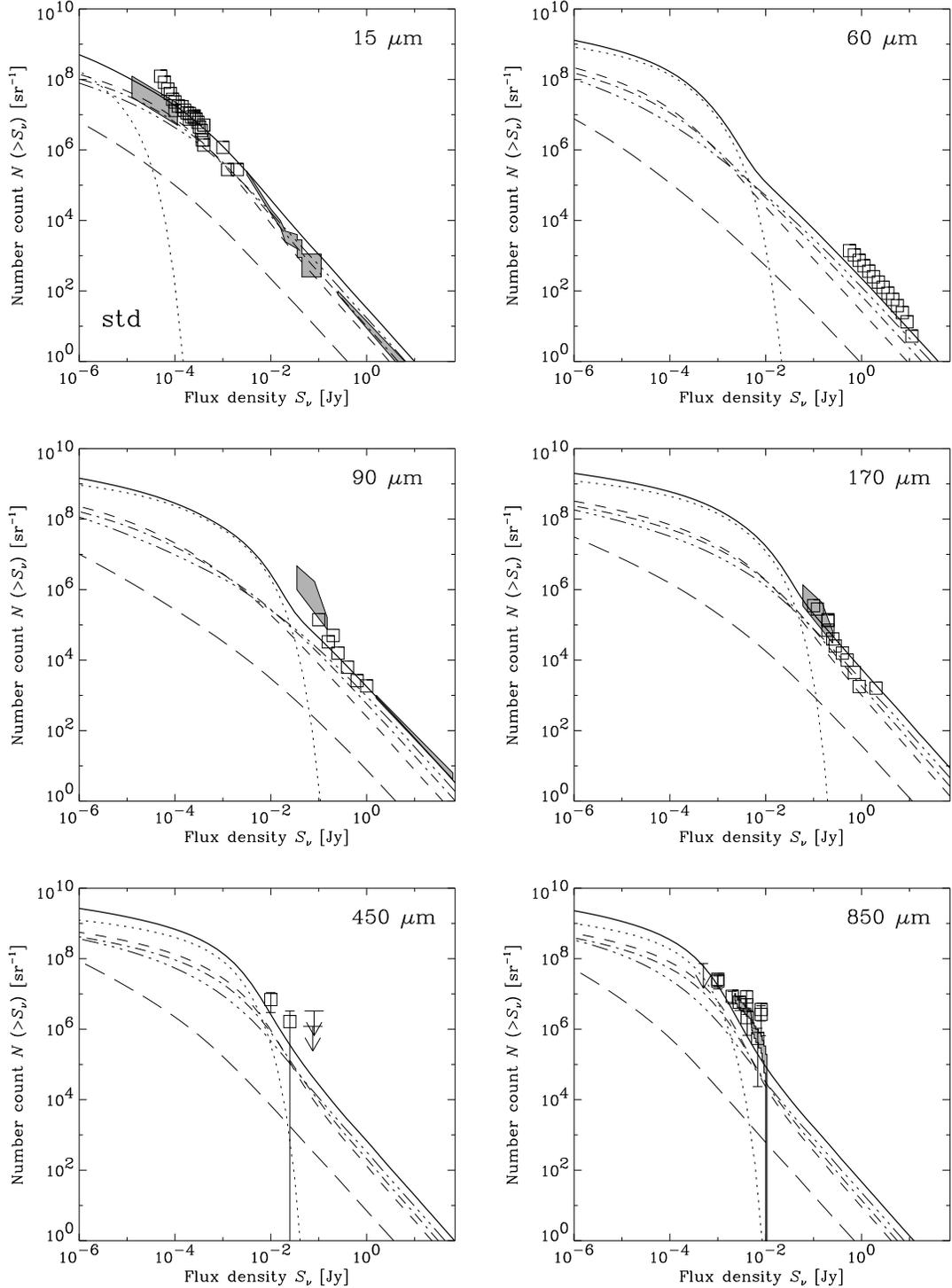}
\caption{
Faint galaxy counts in infrared and submillimeter bands.
The solid line is the prediction by our
baseline model with a cosmological model of $(h, \Omega_0,
\Omega_\Lambda) = (0.7, 0.2, 0.8)$, pure luminosity evolution,
formation redshift of $z_F = 3$, and screen-type distribution of dust.
The solid line is the total of all galaxy types,
while the other five lines are for individual types: E/S0 (dotted),
Sab (short-dashed), Sbc (dot-dashed),  Scd (three-dot-dashed)
and Sdm (long-dashed). 
The data are taken from the following references: 
Rush et al.\ (1993), Oliver et~al.\ (1997), Flores et~al. (1999a, b), 
Clements et~al.~(1999), Aussel et~al.\ (1999), Altieri et~al.~(1999), 
Elbaz et~al.~(1999), and Serjeant et~al.~(2000) 
(15 $\mu$m);
Rowan-Robinson et al.\ (1991) (60~$\mu$m); 
Juvela et al.\ (2000), Matsuhara et~al.\ (2000), Efstathiou et al.~(2000), 
and {\sl IRAS}~100 $\mu$m galaxy counts (90 $\mu$m);
Stickel et~al.\ (1998), Kawara et~al.\ (1998), Puget et~al.\ (1999), 
Juvela et al.~(2000), Matsuhara et~al.\ (2000), and Dole et al.\ (2001)
(170 $\mu$m); 
Smail et~al.~(1997), Barger, Cowie, \& Sanders (1999), and Blain et al.~(2000)
(450 $\mu$m);
and Smail, Ivison, \& Blain (1997), Hughes et al.~(1998), Blain et~al.~(1999), 
Eales et~al.~(1999, 2000), Holland et~al.~(1999), and 
Barger et al.~(1998, 1999) (850 $\mu$m).
}
\label{fig:nc-std}
\end{figure}

Figure~\ref{fig:nc-std} shows predictions of infrared galaxy counts compared 
with available data at 15, 60, 90, 170, 450, and 850 $\mu$m wavebands 
by our baseline model. 
In the baseline model, we assumed a $\Lambda$-dominated flat cosmology
with $(h, \Omega_0, \Omega_\Lambda)$ = (0.7, 0.2, 0.8),
formation redshift of $z_F = 3$ for all galaxy types, 
pure luminosity evolution without number evolution, and screen dust model. 
See figure caption for the references of observed data.
The parameter of overall normalization of dust optical depth,
$A_{V, {\rm MW}}$ is determined here so that the count predictions
fit best the brightest {\sl IRAS}~counts at 15, 60, and 90 $\mu$m bands.
This normalization has already been used in the previous sections.
Our model predicts lower and higher counts than 
the bright counts observed in 15 and 60 $\mu$m, 
respectively, and it may suggest that our modeling of infrared SED might be 
imperfect, but it might also reflect some systematic uncertainties in the 
calibration of the observed flux in these wavebands (see e.g., Mazzei et 
al.\ 2001; Lari et al. 2001; Serjeant et al.\ 2001).\footnote{Calibration 
in the MIR and FIR requires state-of-the-art techniques and
especial scrutiny, and although a lot of efforts by experts have been 
devoted to it, there still remains a significant uncertainty.
We do not go further into the problem here.}
This figure shows contributions from each galaxy type as well as the total 
counts. 
Nearby spiral galaxies are dominant in the brightest counts, while 
contribution from elliptical galaxies rapidly appears when the flux is 
decreased down to a certain value, because of the strong dusty starburst 
of these galaxies at their formation stage. 
Appearance of high-$z$ starburst galaxies revealed by SCUBA in submillimeter 
wavelengths is in good agreement with count predictions of forming elliptical 
galaxies at $z \sim 3$ in our model, and hence it can be interpreted as 
the emergence of forming elliptical galaxies.
The redshift of $z \sim 3$ for these galaxies is also consistent with 
redshift estimates of SCUBA sources (e.g., Dunlop 2001).
On the other hand, our model suggests that relatively weak evolution in the 
15$\mu$m-band counts at $S \lesssim 10^{-2}$Jy  
is not due to elliptical galaxies, but induced by evolution 
of spiral galaxies. 
Strong evolution suggested by {\sl ISO}\ observations (e.g., Efstathiou et 
al.\ 2000; Matsuhara et al.\ 2000) in the 90$\mu$m band may not be explained 
by the emergence of elliptical galaxies in our model. 
However, contribution from such galaxies with smaller $z_F$,
or bulges of spiral galaxies, could explain 
the observed evolution (see Fig. \ref{fig:nc-zF} 
below). (It should be noted that
the bulge and disk of spiral galaxies are modeled as a mixture
in our model.)

\begin{figure}
\plotone{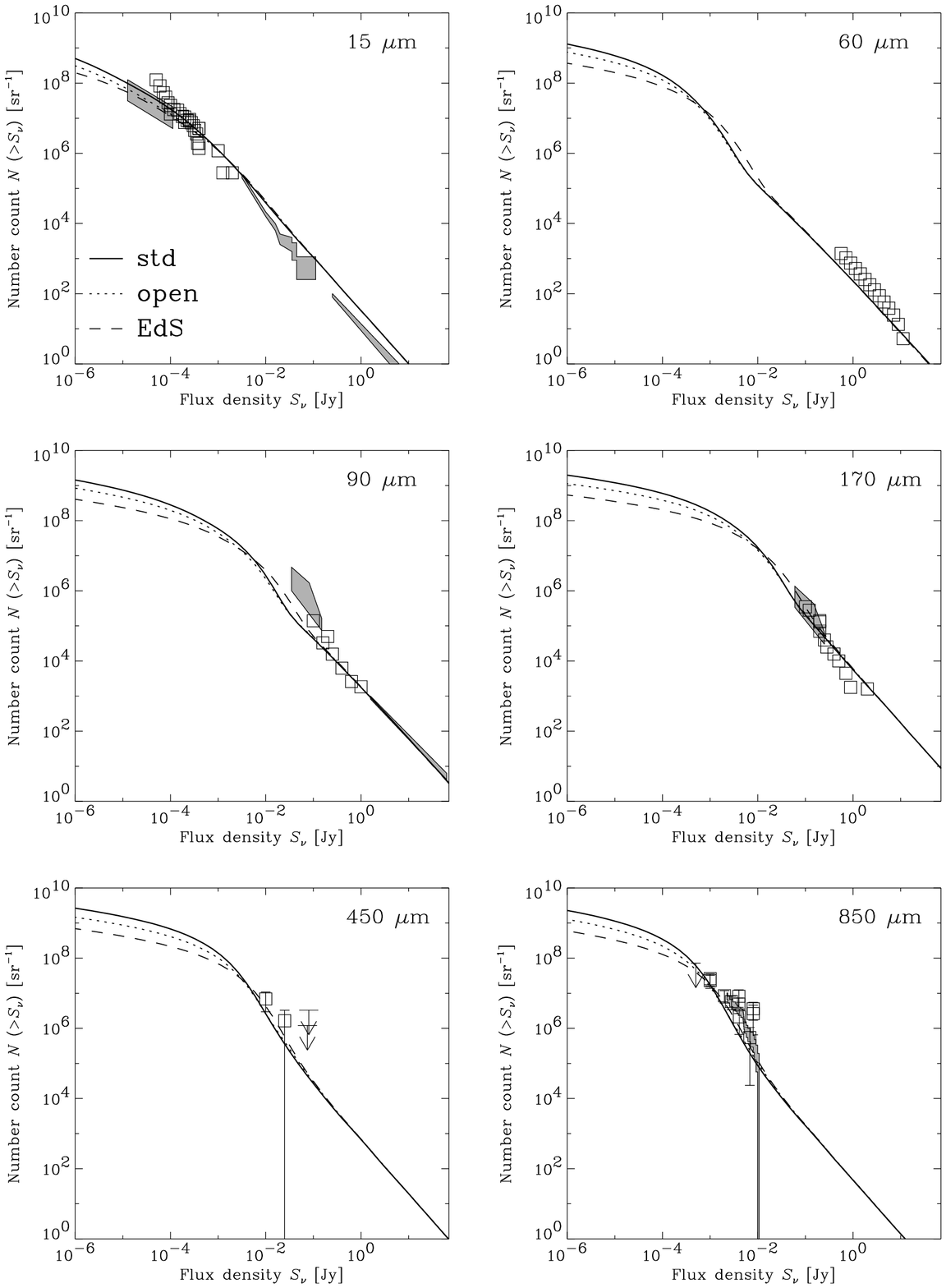}
\caption{The same as Fig. \ref{fig:nc-std}, but for different
cosmological models: the $\Lambda$-dominated flat universe
with $(h, \Omega_0, \Omega_\Lambda)$ = (0.7, 0.2, 0.8) (solid, 
the baseline model shown
in Fig. \ref{fig:nc-std}), an open universe with
(0.7, 0.2, 0.0) (dotted), and an Einstein-de Sitter universe with
(0.5, 1, 0) (dashed).
}
\label{fig:nc-cosm}
\end{figure}

\begin{figure}
\plotone{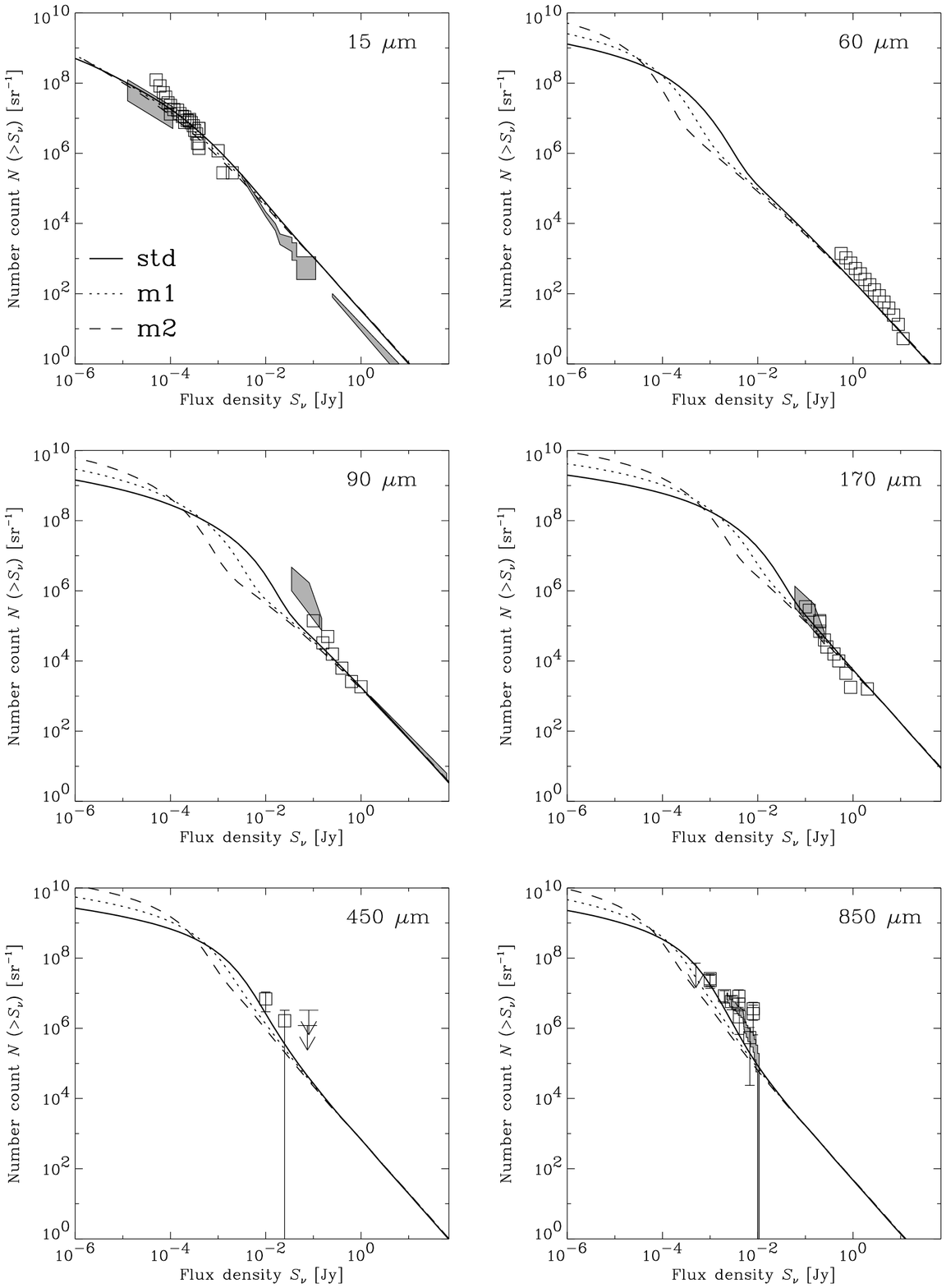}
\caption{The same as Fig. \ref{fig:nc-std}, but for the cases of
number evolution of galaxies: $\eta = 0$ (solid, 
the baseline model without number evolution shown in
Fig. \ref{fig:nc-std}), $\eta = 1$ (dotted), and $\eta = 2$ (dashed).
See text for the modeling of number evolution.
}
\label{fig:nc-num}
\end{figure}

Figures~\ref{fig:nc-cosm}, \ref{fig:nc-zF}, and \ref{fig:nc-num} show the 
count prediction for the total of all galaxy types in cases of different 
cosmological models, formation redshifts, and number evolutions, respectively. 
The dependence on cosmological models becomes important at flux
much fainter than the present sensitivity limits, except for 15$\mu$m.
Therefore the uncertainty in cosmological parameters does not affect
our conclusions significantly. Although there may be some
uncertainties in our modeling of number evolution, it seems that
the counts of SCUBA sources in submillimeter bands are better
explained by primordial elliptical galaxies without number evolution,
rather than the cases of number evolution. 
This may indicate that 
elliptical galaxies have formed by a single starburst at high
redshifts, rather than by many starbursts of smaller building blocks
which would eventually merge into a giant elliptical galaxy 
after the starbursts.
This is consistent with a number of literature about observational 
constraints on number evolution of elliptical galaxies 
(e.g., Totani \& Yoshii 1998; Im et al. 1999; Schade et al. 1999;
Benitez et al. 1999; Broadhurst \& Bouwens 2000; Daddi, Cimatti, \&
Renzini 2000; Totani et al. 2001c).

Figure~\ref{fig:nc-alt} shows some model predictions with different model 
prescriptions: using empirical $L_{\rm IR}$-$T_{\rm dust}$ relation that 
has been used in most of previous studies (dotted line), using slab-type 
dust geometry rather than the screen dust (dashed), and using a different 
prescription for dust opacity by Guiderdoni \& Rocca-Volmerange (1987) as 
discussed in \S \ref{section:model-optical}, rather than assuming a constant 
dust-metal ratio (dot-dashed line). 
The last prescription gives almost the same prediction with our baseline 
model, and hence difference of the two prescriptions for dust/metal ratio is 
not important. 
On the other hand, the empirical $L_{\rm IR}$-$T_{\rm dust}$ relation gives 
quite different prediction especially for elliptical galaxies at the dusty 
starburst phase, because the empirical relation gives considerably lower 
temperature than our baseline model as shown in the previous section. 
Consequently, counts at 90 and 170 $\mu$m are decreased, while those at
450 and 850 $\mu$m are enhanced. The slab dust has the same effect,
although it is weaker than the case of the empirical $L_{\rm IR}$-$T_{\rm 
dust}$ relation.

\subsection{Cosmic Background Radiation}
\label{section:CBR}
\begin{figure}
\plotone{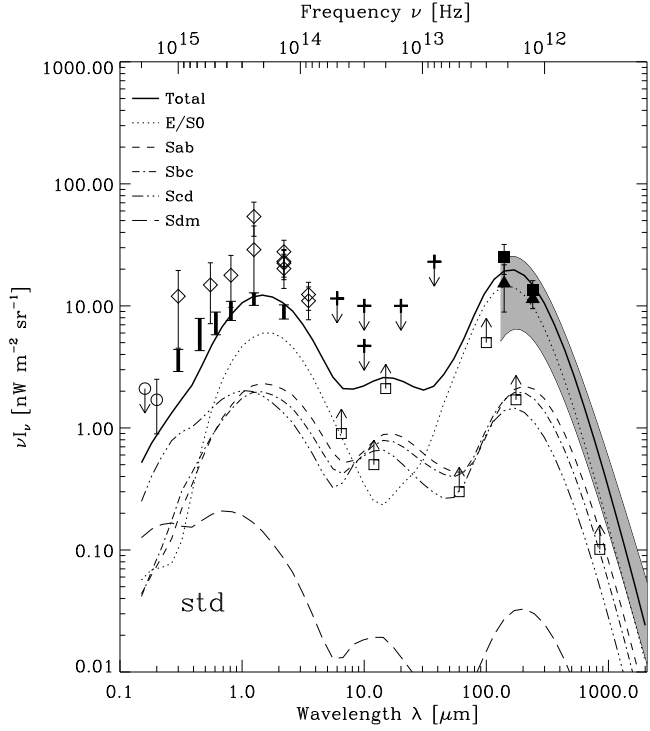}
\caption{Spectrum of the cosmic background radiation (CBR) from
optical to FIR bands. The solid line is the prediction by our
baseline model, and
the other five lines show the contribution from individual types of 
galaxies, as indicated in the figure. 
References for the observational data are as follows;
UV data shown by open circles:
Martin, Hurwitz, \& Bowyer (1991; 1600~\AA) and
Armand et al.\ (1994; 2000~\AA).  
Ranges estimated from integration of optical/NIR galaxy counts:
Totani et al.\ (2001a, thick error bars without symbols). 
Reports for diffuse CBR detections at optical 
and NIR bands are
shown by open diamonds: Bernstein, Freedman, \& Madore (2002; 0.3,
0.55, and 0.8$\mu$m),
Gorjian et al.\ (2000; 2.2 and 3.5$\mu$m), 
Wright \& Reese (2000; 2.2 and 3.5$\mu$m),
Cambr\'{e}sy et al.\ (2001; 1.25 and 2.2$\mu$m), 
and Wright (2001; 1.25 and 2.2$\mu$m).
Lower limits from infrared/submillimeter 
galaxy count integrations shown by open squares: 
Gispert et al.\ (2000; 6.5 $\mu$m), 
Clements et al.\ (1999; 12 $\mu$m), 
Elbaz et al.\ (1999; 15 $\mu$m), 
Lonsdale et al.\ (1990; 60 $\mu$m),
Dwek et al.\ (1998; 90 $\mu$m),
Puget et al.\ (1999; 175 $\mu$m),
Barger et al.\ (1999; 850 $\mu$m).
MIR upper limits from TeV gamma-ray observations shown by 
thick crosses: Biller et al. (1998) and Renault et al.\ (2001).
{\sl COBE}/DIRBE detections at 140 and 240 $\mu$m:
Hauser et al.\ (1998, filled squares) and
Lagache et al.\ (2000, filled triangles).
{\sl COBE}/FIRAS detection at 100 $\mu$m -- 1~mm shown by the shaded area:
Fixsen et al.\ (1998).
}
\label{fig:ebl-std}
\end{figure}

We show our prediction of the CBR in optical and infrared wavelengths
in Fig. \ref{fig:ebl-std}, by our baseline model whose counts have
been presented in Fig. \ref{fig:nc-std}. The reported detections of
CBR, lower limits coming from integration of galaxy counts,
and upper limits on MIR CBR from TeV gamma-ray observations
are also shown (see figure caption for references).
The model is in overall agreement with all available data. 

In optical/NIR bands, the faint-end slope of galaxy counts is very
flat and hence the integration of galaxy counts with extrapolation
down to even fainter magnitude is convergent. Therefore the
integration of galaxy counts in these bands should not be much different
from the true CBR from galaxies.
Since our model is based on a model of galaxy counts in optical/NIR
bands which is in agreement with observations, it is trivial that
our model prediction of optical/NIR CBR is in agreement with 
observational estimates based on integrations of observed galaxy
counts. Such estimates are shown by 
error bars without symbols in Fig. \ref{fig:ebl-std}, in which 
the contribution to CBR by galaxies missed in deep surveys by
various selection effects is taken into account (Totani et al. 2001a).  
However, our model underpredicts
the UV CBR flux compared with the observed count integrations, 
and this suggests that the dust extinction
is too large in the model. In fact, in this paper we have
set the overall normalization of dust, $A_{V, \rm MW} \sim 0.7$, so that the
bright infrared galaxy counts of our model are consistent with 
observed data. This is considerably larger than the typical Galactic
extinction in relatively high latitude, $A_{V, \rm MW} \sim 0.2$,
which was used in our previous papers on optical/NIR counts
(Totani et al. 2001a, b, c). This discrepancy is presumably
coming from the scatter of dust optical depth in galaxies.
In our model we applied a single relation between $B$ luminosity
and galaxy size for the simplicity, but in reality the scatter
along the mean $L_B$-$r_e$ relation should produce significant
scatter of extinction, as discussed
in \S \ref{section:model-optical} and \S \ref{section:local}.
The infrared counts are expected to be
dominated by galaxies with relatively large extinctions, while
the optical/NIR counts, especially $U$ band, would be dominated by
those with less extinctions. This is probably the reason our 
prediction of UV CBR is lower than the observed count integration.
Since the primary interest of this paper is in infrared bands,
we will ignore the discrepancy in UV CBR in the following part of
the paper.

There are several reports of detection of diffuse NIR CBR flux
in the $J$ (1.25$\mu$m), $K$ (2.2$\mu$m) and $L$ (3.5$\mu$m)
bands as shown by diamonds
in Fig. \ref{fig:ebl-std} (Gorjian et al. 2000; Wright \& Reese 2000;
Wright et al. 2001; Matsumoto et al. 2001; Cambr\'esy et al. 2001).
All these reported diffuse NIR CBR fluxes are systematically higher by a 
factor of several than
the integration of galaxy counts in the same bands. Our model
is made to reproduce the observed galaxy counts, and hence our CBR
flux agrees with the galaxy count integrations, and does not agree
with the reported diffuse CBR flux. The contribution from galaxies
missed in deep surveys by selection effects is very unlikely to
reconcile this discrepancy (Totani et al. 2001a).
It may suggest some systematic
uncertainties in diffuse CBR measurements, or the contribution 
to the NIR CBR from yet unknown sources, but it is beyond the scope of
this paper. On the other hand, there is a recent report of detection
of diffuse CBR in optical bands (Bernstein, Freedman, \& Madore 2002).
Although the reported diffuse flux again seems higher than the count
integration estimations by Totani et al. (2001a), they are consistent with 
each other at 1$\sigma$ when all error bars (statistical and
systematic of the diffuse measurements, and contribution from faint galaxies
missed by deep surveys) are taken into account. 

\begin{figure}
\plotone{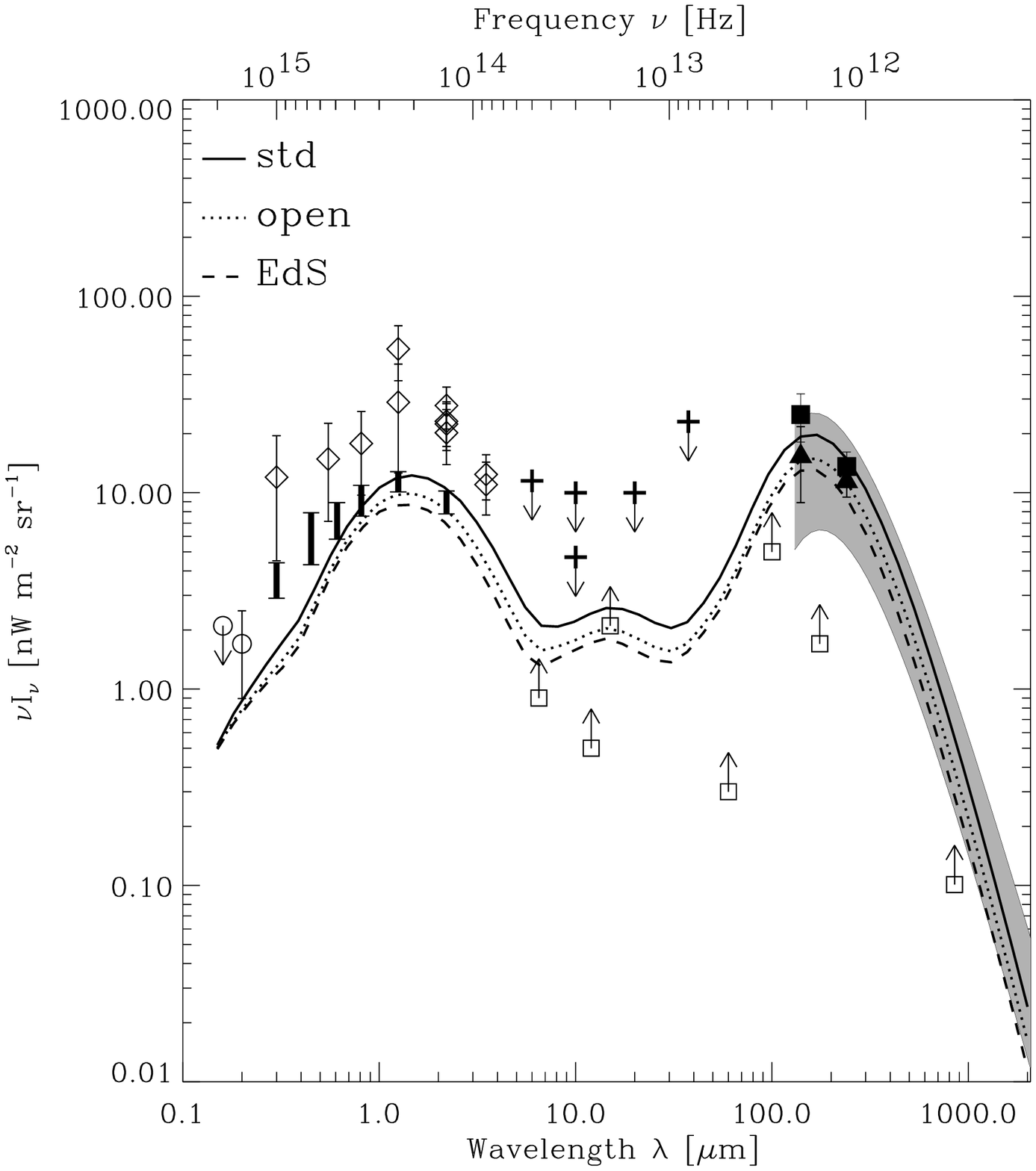}
\caption{The same as Fig. \ref{fig:ebl-std}, but for different
cosmological models: the $\Lambda$-dominated flat universe
with $(h, \Omega_0, \Omega_\Lambda)$ = (0.7, 0.2, 0.8) (solid, 
the baseline model shown
in Fig. \ref{fig:ebl-std}), an open universe with
(0.7, 0.2, 0.0) (dotted), and an Einstein-de Sitter universe with
(0.5, 1, 0) (dashed).
}
\label{fig:ebl-cosm}
\end{figure}

\begin{figure}
\plotone{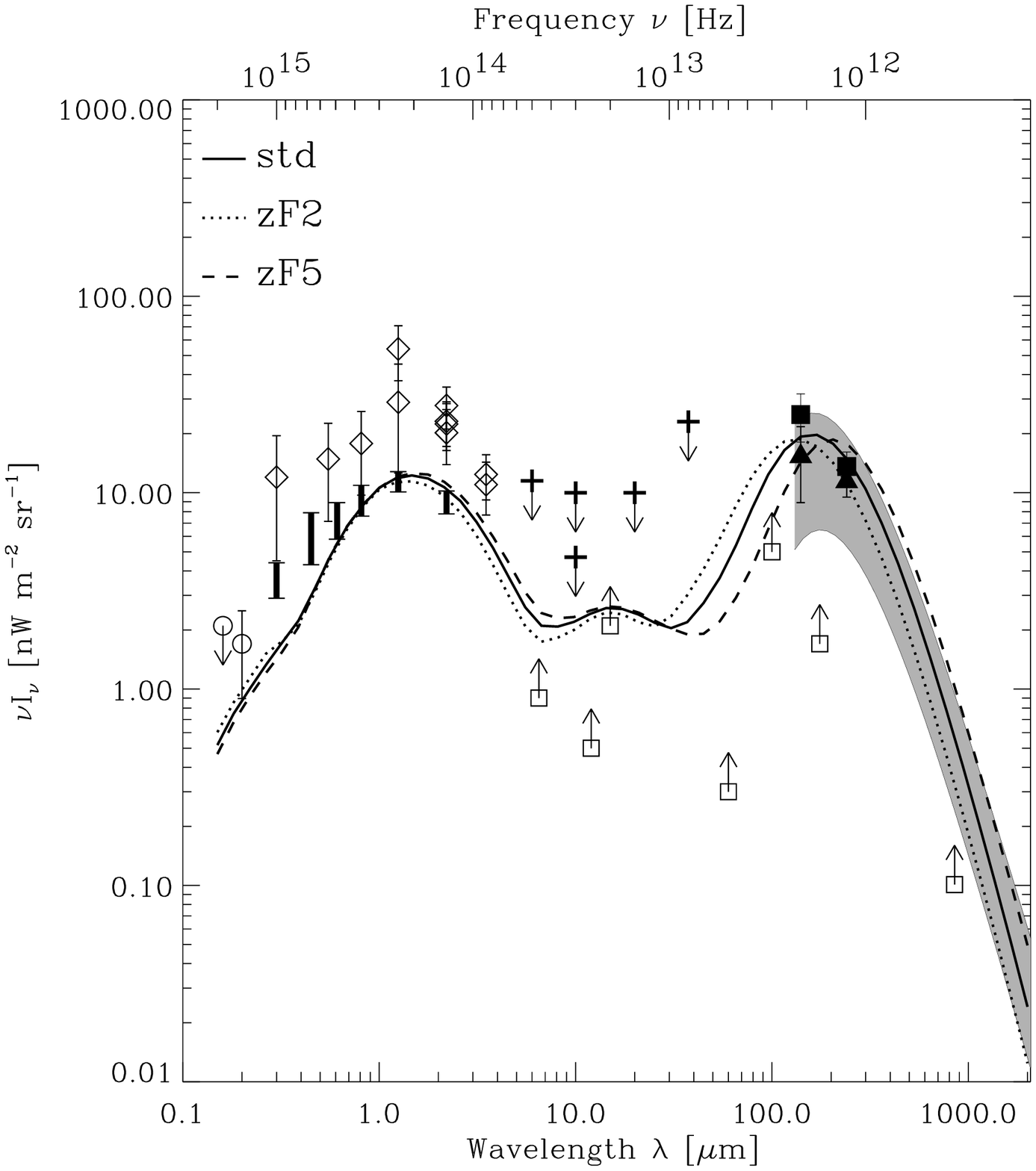}
\caption{The same as Fig. \ref{fig:ebl-std}, but for different
formation redshift of galaxies: $z_F = 2$ (dotted), $z_F = 3$
(solid, the baseline model shown in Fig. \ref{fig:ebl-std}), and
$z_F = 5$ (dashed).
}
\label{fig:ebl-zF}
\end{figure}

\begin{figure}
\plotone{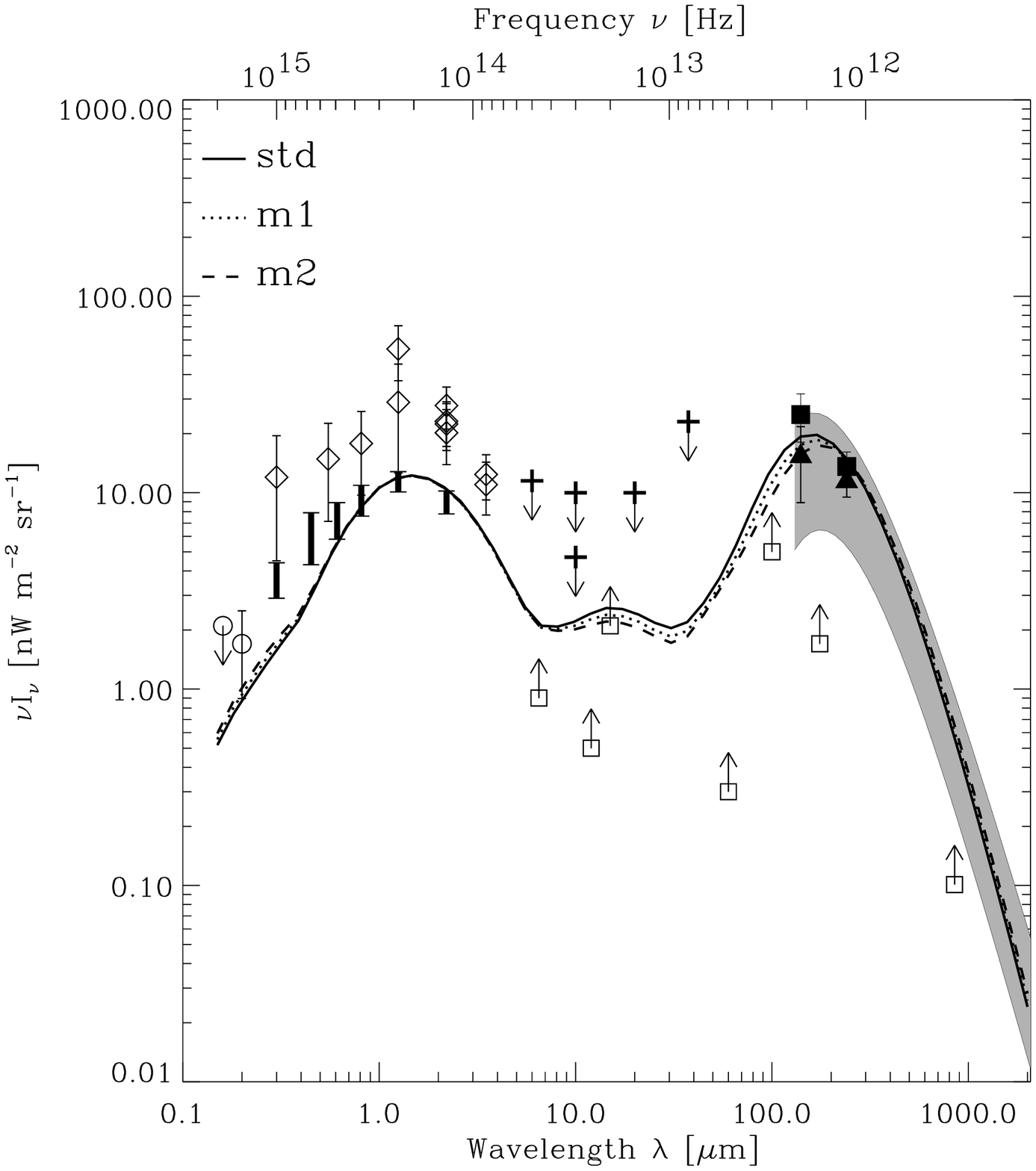}
\caption{The same as Fig. \ref{fig:ebl-std}, but for the cases of
number evolution of galaxies: $\eta = 0$ (solid, 
the baseline model without number evolution shown in
Fig. \ref{fig:ebl-std}), $\eta = 1$ (dotted), and $\eta = 2$ (dashed).
See text for the modeling of number evolution.
}
\label{fig:ebl-num}
\end{figure}

Figures \ref{fig:ebl-cosm}, \ref{fig:ebl-zF}, and \ref{fig:ebl-num}
show the CBR prediction for the total of all galaxy types
in cases of different cosmological models, formation redshifts, and
number evolutions, respectively. These figures show that
the dependence on cosmological models or number evolution
is not large compared with observational constraints, and it
does not affect our conclusions significantly.
On the other hand, the formation redshift can 
be constrained better since the sub-mm CBR flux rapidly increase 
with increasing $z_F$ due to the redshift effect on the
FIR peak of dust emission SED, while there is
an observational constraint by COBE/FIRAS.
Our model suggests that the major formation epoch of elliptical galaxies 
should not be larger than $z_F \sim 5$.

\begin{figure}
\plotone{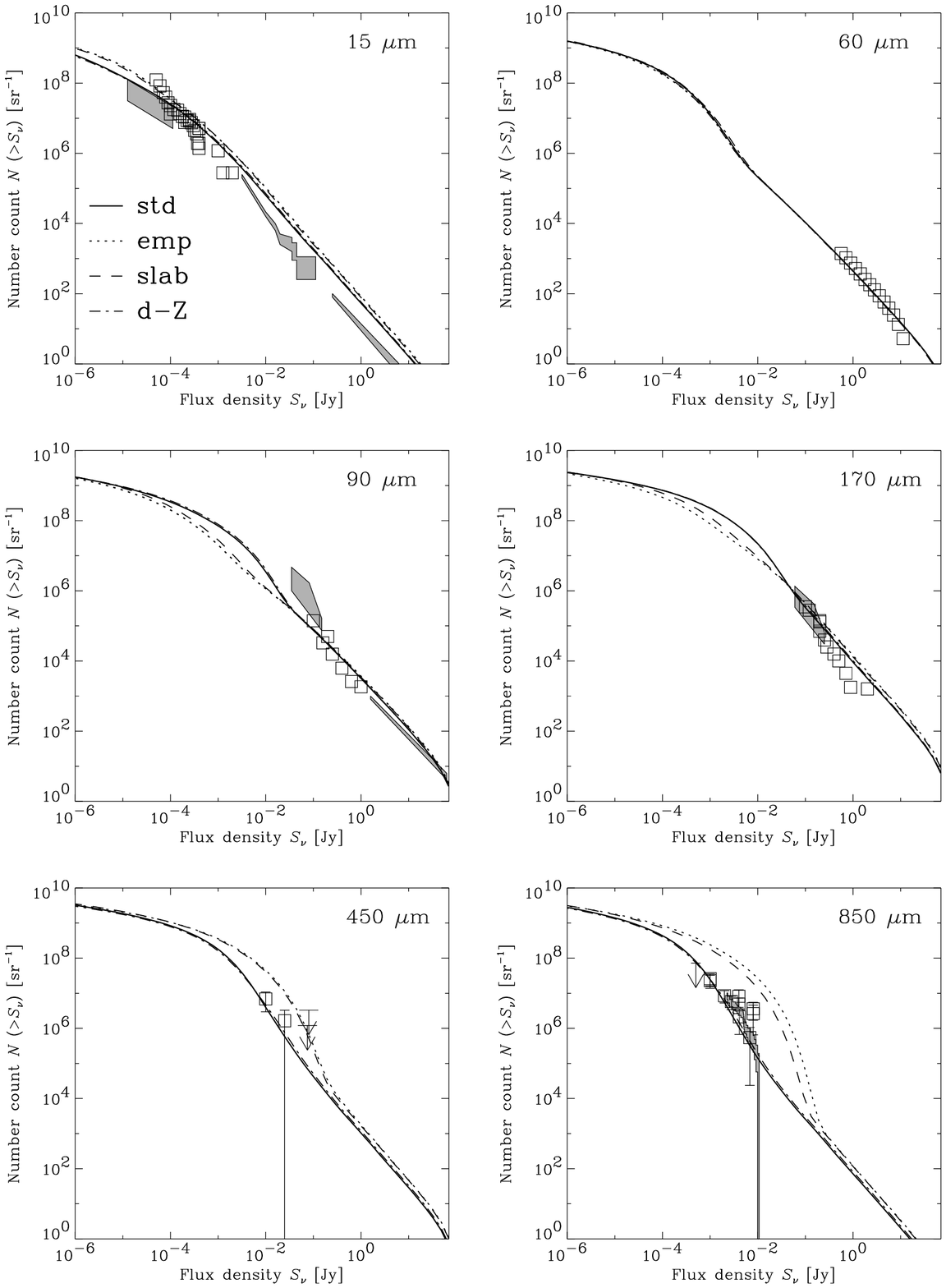}
\caption{The same as Fig. \ref{fig:nc-std}, but with different
model prescriptions from the baseline model (solid line, and shown
in Fig.\ref{fig:nc-std}): empirical $L_{\rm IR}$-$T_{\rm dust}$ 
relation (dotted), slab-type dust (dashed), and non-proportional
dust/metal ratio (dot-dashed).
}
\label{fig:nc-alt}
\end{figure}

\begin{figure}
\plotone{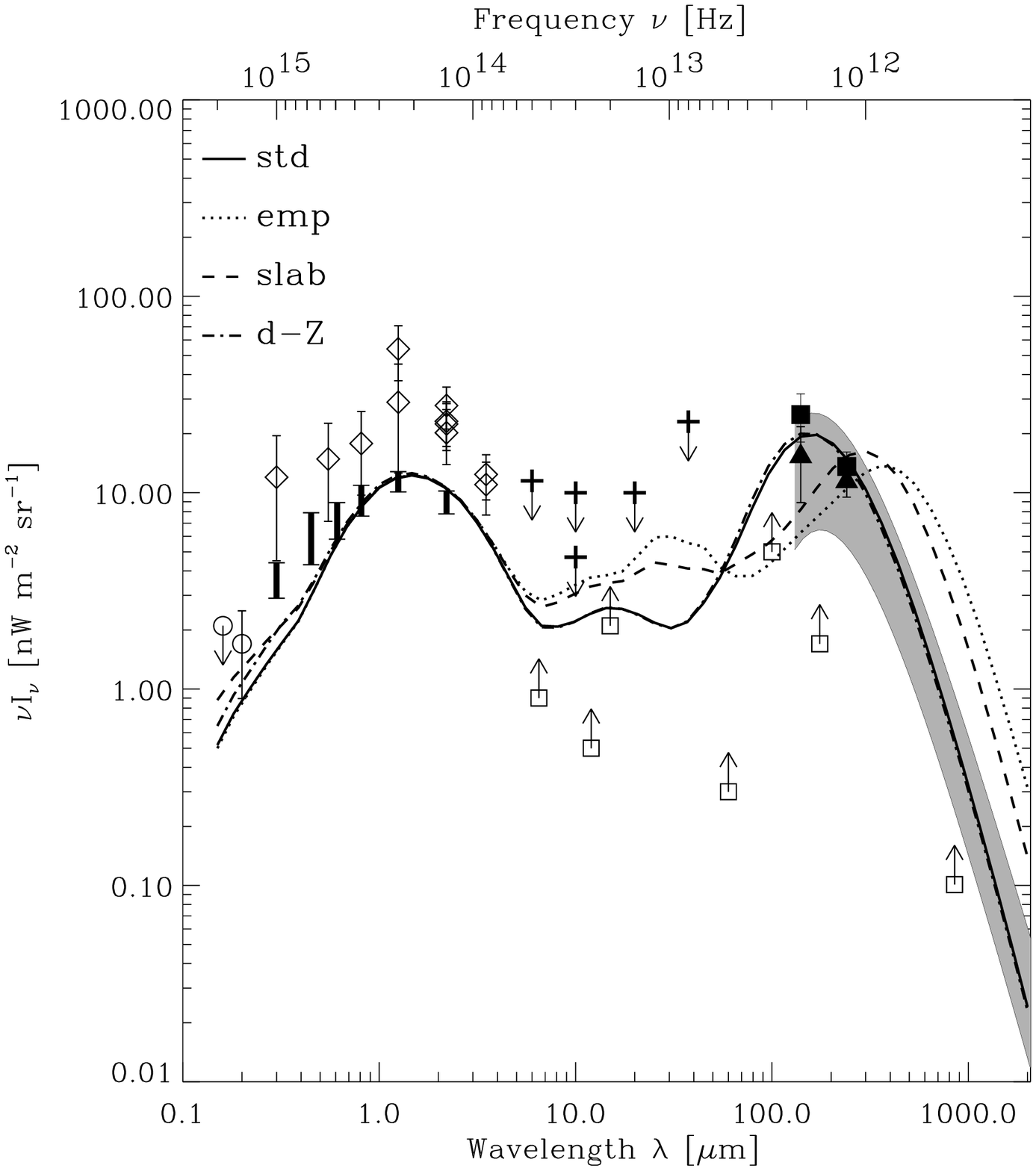}
\caption{The same as Fig. \ref{fig:ebl-std}, but with different
model prescriptions from the baseline model (solid line, and shown
in Fig.\ref{fig:ebl-std}): empirical $L_{\rm IR}$-$T_{\rm dust}$ 
relation (dotted), slab-type dust (dashed), and non-proportional
dust/metal ratio (dot-dashed).
}
\label{fig:ebl-alt}
\end{figure}

Figure \ref{fig:ebl-alt} shows some model
predictions with different model prescriptions, with the same models
and line markings with those in Fig. \ref{fig:nc-alt}.
There is again a remarkable difference between the baseline model and
the model using the empirical $L_{\rm IR}$-$T_{\rm dust}$ relation;
the latter model seriously overpredicts the sub-mm CBR flux compared with 
the detection of the {\sl COBE}/FIRAS. 
When the empirical $L_{\rm IR}$-$T_{\rm dust}$ relation is used, 
temperature of starbursting elliptical galaxies does not become as high as 
our baseline model, and hence the redshift effect significantly pushes FIR 
peak of CBR toward longer wavelengths. 
In fact, because of this reason, most of the previous papers published 
recently have claimed that too intense starbursts well beyond $z \sim 1$ are
disfavored by the FIRAS data.
However, our model suggests, based on a physical argument, that
high-$z$ starbursts of forming elliptical galaxies should have
much higher temperature than previously expected from the empirical
$L_{\rm IR}$-$T_{\rm dust}$ relation, and hence 
formation of elliptical galaxies at high redshifts ($z_F \sim 3$) with dusty
starbursts is allowed, or even favored, from the data.
This is consistent with the standard picture of elliptical
galaxy formation (Larson 1974; Arimoto \& Yoshii 1987), i.e., 
intense initial starburst at high redshift followed by passive evolution
to the present.
The screen dust model seems much better than the slab-type model,
as shown in Fig. \ref{fig:nc-alt} and \ref{fig:ebl-alt}. 
This has an interesting implication
for the optical/NIR colors of primordial elliptical galaxies in dusty 
starburst phase, and we will discuss it in the next section.

\section{Implications and Discussion}
\label{section:discussion}
\subsection{Forming elliptical galaxies in optical/NIR bands: HEROs}
We have shown that the screen-like dust extinction gives quantitatively
much better results both for the infrared counts and CBR
than the slab-like dust, if elliptical
galaxies formed by starbursts at high redshifts of $z_F \gtrsim$ 2
as generally believed. 
This is consistent with the properties of nearby starburst galaxies showing 
strong reddening which cannot be explained only by the slab-like dust. 
It also seems reasonable that at least a part of dust particles is blown out 
by strong galactic wind expected in the formation stage of elliptical 
galaxies, to have a distribution similar to a screen. 
An implication of this result is that forming elliptical galaxies should 
show very red colors when observed in optical/NIR wavelengths. 
When dust opacity becomes much greater than unity, colors of galaxies could 
infinitely increase by screen dust, while colors should reach an asymptote 
in the case of slab dust since emission from stars located at the outermost 
region becomes dominant. 

In fact, such objects having
unusually NIR colors of $J-K \gtrsim$ 3--4 were discovered
recently in the Subaru Deep Field (Maihara et al. 2001;
Totani et al. 2001b). These colors are even redder than those of
typical extremely red objects (EROs) selected by optical-NIR colors,
and cannot be explained by
passively evolving dust-free elliptical galaxies.
Totani et al. (2001b) argued that the
colors, magnitudes, and counts of these
enigmatic objects can be best explained by forming elliptical
galaxies reddened by screen-like dust. It was also found that
the surface number density of such hyper-extremely red objects (HEROs) is
roughly the same with that of high-$z$ starbursts discovered
by SCUBA in sub-mm bands. In the count model of this paper,
a significant part of the SCUBA sources can be explained by
the forming elliptical galaxies, and hence our result here
quantitatively strengthens the suggestion of Totani et al. (2001b)
that the SCUBA sources and HEROs have the same origin of
forming primordial elliptical galaxies in the dusty starburst phase.

\subsection{The cosmic star formation history}
\begin{figure}
\plotone{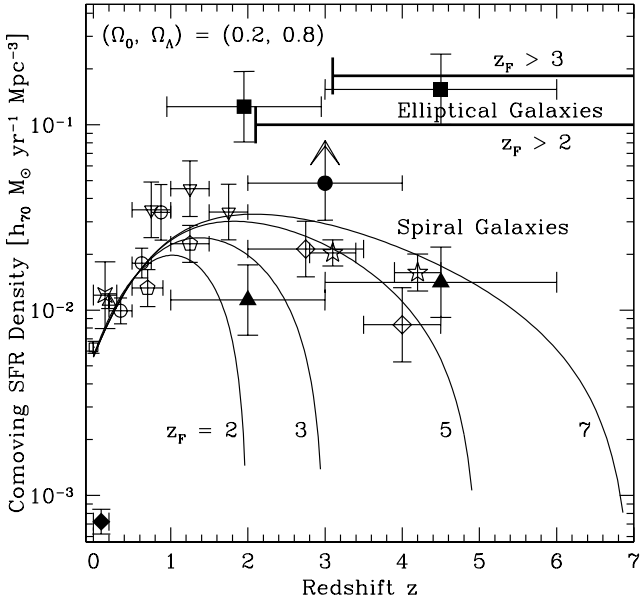}
\caption{Cosmic history of global star formation rate. The solid curves
are the prediction of our baseline model for spiral galaxies (i.e., except for
elliptical galaxies), with several values of $z_F$. The two horizontal
solid lines are mean SFR of elliptical galaxies when they are assumed
to be formed before $z_F$= 2 or 3, respectively. The observed data
for dust-uncorrected SFR from H$\alpha$ or UV luminosity density 
(i.e., unabsorbed SFR) are shown by open symbols:
Gallego et al. (1995, square), Tresse \& Maddox (1998,
triangle), Treyer et al. (1998. four-edged star), 
Lilly et al. (1996, open circles), Connolly et al. (1997, upside-down
triangles),  Madau, Pozzetti, \& Dickinson (1998, diamonds), 
Steidel et al. (1999, five-edged stars), Cowie, Songaila, \& Barger
(1999, pentagons).
The SFR density inferred from submillimeter observations (i.e., hidden SFR)
are shown by filled symbols:
Hughes et al. (1998, circle); Barger, et al. (2000)
before (triangles) and after (squares) the completeness corrections.
The local hidden SFR density from ULIRGs is shown by the filled diamond
(Barger et al. 2000). These observed data have been corrected
for the cosmological parameters chosen here
[$(h, \Omega_0, \Omega_\Lambda) = (0.7, 0.2, 0.8)$].
}
\label{fig:csfh}
\end{figure}

Figure \ref{fig:csfh} shows the cosmic history of global star
formation rate as a function of redshift, for our model as well as
observational estimates coming from H$\alpha$,
UV, and submillimeter luminosity
density evolution. The elliptical galaxies in our model have a
strong peak of SFR at their formation redshift, while the formation
redshift should be distributed in some redshift range in reality.
Therefore we plot the mean SFR for elliptical galaxies when they
are assumed to be formed before $z$ = 2 or 3, as indicated
by horizontal lines.

Elliptical galaxies in our model have much higher global SFR at 
$z \gtrsim 3$ than the peak at $z \sim 1$ suggested by the optical observations
(Lilly et al. 1996; Connolly et al. 1997; Madau, Pozzetti, \& Dickinson 1998;
Steidel et al. 1999; Cowie, Songaila, \& Barger 1999; Thompson et al. 2001), 
and the model is still consistent with the FIR 
background detected by {\sl COBE}/FIRAS. This is in very sharp
contrast to a number of previous studies 
(e.g., Gispert, Lagache, \& Puget 2000; Takeuchi et al. 2001a; 
Franceschini et al. 2001) which claimed that such high SFR beyond $z\sim1$ is 
disfavored from the {\sl COBE}/FIRAS data. 
This is because the dust temperature of forming elliptical galaxies is much 
higher than expected from the local $L_{\rm IR}$-$T_{\rm dust}$ relation,
as explained in \S \ref{section:CBR}.
The SFR expected from elliptical galaxies is close to those inferred
from galaxies detected in submillimeter observations 
(Hughes et al. 1998; Barger, Cowie, \& Richards 2000), and this
again suggests that a considerable part of forming elliptical galaxies
has been detected by submillimeter observations. [Note, however,
that the higher points of Barger et al. (2000) are corrected by a factor
of $\sim$ 11 for incompleteness.] On the other hand,
SFR expected from spiral galaxies is in reasonable agreement with
those obtained by optical (i.e., UV in the restframe) observations.
Therefore a natural interpretation of these results is that 
the cosmic star formation history inferred from UV light is
that for steady or modest star formation in disks of spiral galaxies,
while starbursts in elliptical or spheroidal components of galaxies
emit their energy mostly in the form of dust emission observed in
submillimeter wavelengths.

\subsection{On the origin of optical size-luminosity relation}
The key of our success to reproduce the scaling  properties of local
infrared galaxies (i.e., optical/infrared luminosity ratio, 
$L_{\rm IR}$-$T_{\rm dust}$ relation, and infrared luminosity
function) is the relation between the optical luminosity and
size of galaxies, which is used as an input in our model.
Therefore the readers may wonder what is the origin of this 
relation. In fact, this scaling relation can be explained by
the framework of the standard structure formation theory in the CDM
universe. One can predict the virial size of cosmological objects
when the mass and formation redshift are specified. Then the observed
scaling relation between $L_B$ and $r_e$ is easily reproduced 
when some reasonable assumptions are made such as a roughly constant
mass-to-luminosity ratio, proportionality between the virial radius
and disk effective radius (equivalent to the constant angular momentum), 
and formation of disks at $z \lesssim 1$ (e.g., Mao \& Mo 1998).
The scatter along the mean relation can also be reproduced by
the dispersion of angular momentum of dark haloes.
This means that the local scaling relations of infrared galaxies
also have their origin in the structure formation in the CDM universe.
This also explains 
the success of a semi-analytic model of Granato et al.\ (2000)
to reproduce the local infrared properties.

\subsection{Implications for the TeV gamma-ray observations}
\label{section:TeV}
TeV gamma-ray observations provide a unique constraint on
CBR because very high energy
gamma-rays beyond $\sim$TeV are absorbed by interaction with
the infrared CBR photons to produce electron-positron pairs in
intergalactic field (e.g., Gould \& Schr\'eder 1967;
Stecker \& de Jager 1997, 1998; Salamon \&
Stecker 1998; Primack et al. 1999). Gamma-rays with energy
$\epsilon_\gamma$ mainly interact with infrared photons with
wavelength $\lambda \sim h_P \epsilon_\gamma / (4 m_e^2 c^3)
\sim 1.24 \epsilon_{\gamma, \rm TeV} \ \rm \mu m$, where
$h_P$ is the Planck constant and $\epsilon_{\gamma, \rm TeV}
= \epsilon_\gamma$/TeV. Therefore detections of gamma-rays 
at $\sim$ 10TeV for TeV blazars Mrk 421 ($z = 0.031)$ 
and Mrk 501 $(z = 0.034$) (Krennrich, et al.\ 1999; Aharonian et al.\ 1999) 
give constraints on the MIR CBR at about 10--20 $\mu$m. 
The thick crosses plotted in Fig.~\ref{fig:ebl-std} are taken from 
Biller et al.\ (1998) and Renault et al.\ (2001), and our baseline model is 
consistent with this limit. 
The CBR flux at 15$\mu$m in our baseline model is 2.5nW/m$^2$/sr, and the 
lower bound set by the count integration in this band accounts for about 
80~\% of the total CBR of the model.

It should be noted that the model using the empirical 
$L_{\rm IR}$-$T_{\rm dust}$ relation predicts higher MIR CBR flux,
while the FIR peak
is lower than the baseline model (Fig. \ref{fig:ebl-alt}).
This difference originates from the dependence of MIR/FIR flux ratio
of the model infrared SEDs on the dust temperature. 
As can be seen in Fig.~\ref{fig:SED-schem}, MIR/FIR flux ratio decreases with 
increasing dust temperature, due to the diminishing component of small grain
dust and PAH features. This trend is physically reasonable as
explained in \S \ref{section:SED}.
The difference of our baseline model and that using
the empirical $L_{\rm IR}$-$T_{\rm dust}$ relation is that
the dust temperature of starbursting elliptical galaxies in the former
is much higher than in the latter. Therefore, in the baseline 
model, the strong thermal
dust emission from forming elliptical galaxies has a largest contribution to 
the peak of FIR CBR, while it hardly contributes to the MIR CBR 
(Fig.~\ref{fig:ebl-std}). 
This is why we could 
reproduce simultaneously the high FIR peak and low MIR CBR flux which are
fully consistent with observations of {\sl COBE}\ and TeV gamma-rays.

On the other hand, the dust temperature does not become as high as 
our baseline model in models employing the empirical $L_{\rm
IR}$-$T_{\rm dust}$ relation, and hence MIR/FIR flux ratio of CBR
is not low enough to match both TeV limits in MIR and 
the {\sl COBE}\ data in FIR. 
This {\it prima facie} conflict has been noticed
in previous work based on the empirical $L_{\rm IR}$-$T_{\rm dust}$
relations of local infrared galaxies (e.g., Stecker 2000), and 
several possibilities have been proposed to solve this conflict, including 
new physics such as a possibility of the broken Lorentz invariance. 
However, our model suggests that
the MIR and FIR CBR are made by completely different populations
of galaxies: low-redshift spiral galaxies for MIR while high-redshift
starbursting elliptical galaxies for FIR, as seen in Fig. 
\ref{fig:ebl-std}.  Thus in our model the
conflict is naturally resolved.

\begin{figure}
\plotone{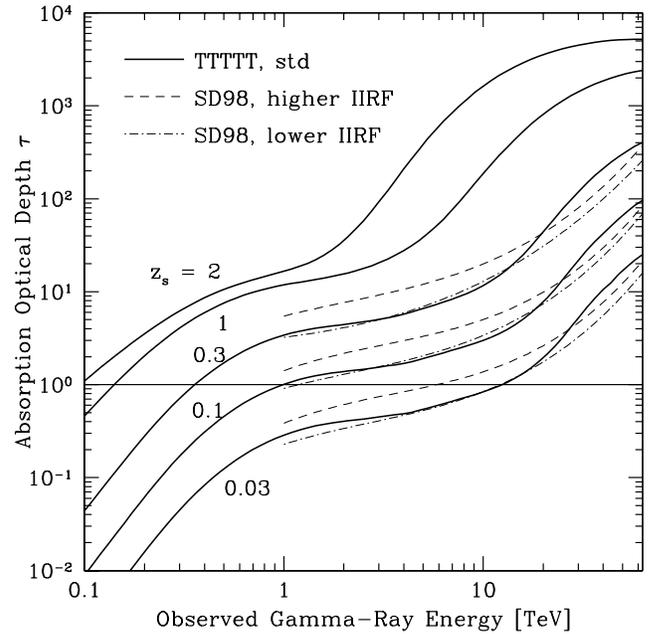}
\caption{The intergalactic optical depth of very high energy
gamma-rays to the absorption by interaction with optical/infrared
cosmic background radiation, as a function of the source redshift ($z_s$)
and the gamma-ray energy observed at $z=0$. The solid lines are
the calculation based on our baseline model, with the source redshifts
indicated in the figure (indicated as TTTTT from the initials of the
authors).  For comparison, 
calculations with `higher IIRF SED' (dashed line) and `lower IIRF SED'
(dot-dashed line) by Stecker \& de Jager (1998) are plotted for
$z_s$ = 0.03, 0.1, and 0.3 (from the bottom to the top).
}
\label{fig:tau}
\end{figure}

We calculated the optical depth of TeV gamma-rays as a function of
source redshift ($z_s$) and gamma-ray energy observed at $z=0$, based on the
formulations given in Salamon \& Stecker (1998). The result is shown
in Fig. \ref{fig:tau}, along with the previous calculations 
of Stecker \& de Jager (1998, hereafter SD98) based on the
empirical $L_{\rm IR}$-$T_{\rm dust}$ relation. 
The baseline model is used for the 
flux and its evolution of the CBR from optical to FIR.
As expected, the optical depth of our model (indicated as TTTTT
in the figure from the initials of the authors) is relatively smaller
at gamma-ray energies of $\sim$ 1--10 TeV where the main target
photons are in the MIR band, while it becomes larger for higher
energy gamma-rays interacting with FIR photons, when compared with the
SD98 calculation. The different two models might be discriminated
by more accurate measurement of the TeV spectrum of Mrk 421 and/or 501
in future observations.
Following SD98, we give the result of polynomial
fits to the optical depth $\tau(z_s, \epsilon_\gamma)$, as
\begin{equation}
\tau(z_s, \epsilon_\gamma) = \sum_{0 \leq i \leq 3} \ \sum_{0 \leq j \leq 2}
a_{ij} (\log_{10} \epsilon_{\gamma, \rm TeV})^i (\log_{10} z_s)^j \ ,
\end{equation}
where the values of $a_{ij}$ are presented in Table
\ref{table:TeVopt}. This fitting formula gives a reasonable
approximation ($<$ 10\% at $ z_s = 0.03$ and $1 <
\epsilon_{\gamma, \rm TeV} < 20$, and $\lesssim$ 50\%
within ranges of $0.03 < z_s < 0.3$ and
$0.1 < \epsilon_{\gamma, \rm TeV} < 50$).
More accurate table of the optical depth 
with wider ranges of $z_s$ and $\epsilon_\gamma$ is available
on request to the authors.

\section{Summary and Conclusions}
\label{section:conclusions}
In this paper we developed a new model of counts and cosmic background
radiation (CBR) of infrared galaxies observed
by emission from heated interstellar dust,
by extending a model for optical/NIR galaxies (Totani \& Yoshii 2000).
Five morphological types of galaxies (E/S0, Sab, Sbc, Scd, and Sdm)
are taken into account and their number densities are normalized
by type-dependent $B$-band luminosity function at the local universe.
Their luminosity evolution is traced backwards in time based on 
star formation histories inferred from the present-day optical/NIR SEDs and 
chemical properties. 
Formation epoch of galaxies is a parameter for which we tried a redshift 
range of $2 \leq z_F \leq 5$. 
Pure luminosity evolution without number evolution is assumed in our 
baseline model, but some number evolution is also tested in a 
phenomenological way. 
The model has already been compared comprehensively with the counts and 
redshift distributions of galaxies observed in optical and NIR bands, 
and found to be in reasonable agreement with the data
(Totani \& Yoshii 2000; Totani et al. 2001c). 

Relatively rare populations of AGNs and 
ULIRGs seen in the local universe, 
whose contribution to counts and CBR is expected to be small, are
not included in our model, and we tried to explain the bulk of infrared data
by normal galaxy populations at the local universe and their ancestors
at high redshifts. On the other hand, our analysis strongly indicates
that the primordial elliptical galaxies are very similar to
dusty starburst galaxies or ULIRGs with very high dust temperature.

There are two important new characteristics in our model that are
different from previous models of infrared galaxies: (1) mass scale
dependence of dust extinction is introduced by the observed
size-luminosity relation of optical galaxies, and (2) dust temperature
is determined by physical consideration of energy balance, rather than
using the empirical relation between the total infrared luminosity 
($L_{\rm IR}$)
and characteristic dust temperature ($T_{\rm dust}$) of local galaxies
that has been used in a number of previous models. 
As a result, the local properties of infrared galaxies, such as 
optical/infrared luminosity ratios, correlation between infrared luminosity 
and dust temperature, and infrared luminosity function, are outputs that 
should be compared with observed data. 
Indeed we found that our model quantitatively reproduces the observed 
infrared properties at the local universe. 
The key to understand these scaling properties is the size-luminosity
relation of galaxies; surface brightness and dust column density
increase with increasing optical galaxy luminosity, and hence more 
massive galaxies should be more dusty. This gives a quantitative
explanation for the observed correlation between
optical and infrared luminosities. 
Furthermore, massive galaxies should emit more energy as dust emission
per unit mass of dust, and hence the energy balance inevitably
results in higher temperature for larger galaxies, as observed
for local infrared galaxies. The scatter along the mean
$L_{B}$-$r_e$ relation is comparable with those in 
$L_{\rm IR}$-$L_{B}$ and $L_{\rm IR}$-$T_{\rm dust}$ relations.
These effects result in much faster
increase of 60$\mu$m luminosity when $L_B$ is increased, and 
giving an explanation for the much broader shape of 60$\mu$m
luminosity function than the Schechter function of the optical
luminosity function.

Then we predicted faint source counts and CBR composed of high-$z$ galaxies. 
Our baseline model assumes a cosmological model with ($h, \Omega_0, 
\Omega_\Lambda$) = (0.7, 0.2, 0.8), pure luminosity evolution after formation 
at $z_F = 3$, and screen distribution of dust.  
We found that this baseline model is in reasonable agreement with all 
available data of galaxy counts in six wavebands (15, 60, 90, 170, 450, and 
850$\mu$m) and CBR. 
Therefore our model, based only 
on present-day normal galaxy populations and their evolution,
reasonably fits to all available data
from optical to submillimeter wavebands, though some modest number
evolution may be required for late-type galaxies. The high-$z$ starburst
galaxies discovered by SCUBA are quantitatively well explained
in our model by the emergence of starbursts at the formation of
present-day elliptical galaxies at $z \sim 3$.

We also tested that slab-type dust distribution as well as the
screen dust used in the baseline model, and found that the screen
model gives much better fit to the observed data of the local infrared
luminosity function, galaxy counts, and CBR. Although a pure screen
distribution may seem unlikely, it is rather reasonable to expect
that a part of dust particles behave like an effective screen,
because of stellar/galactic wind or inhomogeneity of interstellar
medium. If it is the case, and the dust opacity to optical/NIR is
much larger than the unity,
the screen model should be better than the slab-type prescription.
It is also supported by extremely 
red colors of local starburst galaxies
(Calzetti, Kinney, \& Storchi-Bergmann 1994) or recently discovered
hyper extremely red objects (Totani et al. 2001b), which cannot be
explained simply by slab-type dust. Therefore we conclude that
screen dust is a better phenomenological description than slab dust
at least in a study of this kind.

The most drastic difference
of our model from previous ones is that the dust temperature
of starbursting elliptical galaxies is predicted to be much higher
($\sim$ 40--80K) than that extrapolated by the empirical
$L_{\rm IR}$-$T_{\rm dust}$ relation of local infrared galaxies.
This is because starbursting elliptical galaxies should emit much
larger amount of energy as dust emission {\it per unit dust mass},
than local galaxy populations. On the other hand, such high
temperature is similar to those found in local ULIRGs or high-$z$
dust starbursts observed in sub-mm bands. Thus, our result gives
a further support to an idea that 
the progenitor of present-day elliptical galaxies or bulges are
dusty starbursts.

There is an important implication
for the cosmic star formation history, which is very different from
previous results. A number of papers based on the empirical
$L_{\rm IR}$-$T_{\rm dust}$ relation claimed that
cosmic star formation rate beyond $z \gtrsim$ 1 must turn over
and keep constant or decline, otherwise it would produce too much
submillimeter CBR compared with the {\sl COBE}/FIRAS data, by the
redshifted dust emission from high-$z$ galaxies
(e.g., Gispert, Lagache, \& Puget 2000; Takeuchi et al.\ 2001a;
Franceschini et al.\ 2001).
However, in our model the dust temperature of forming 
elliptical galaxies is much higher than in the previous models.
As a result, although our model assumes very 
strong starbursts at $z_F \sim$ 3 in primordial elliptical galaxies
and the cosmic SFR at $z_F \gtrsim$ 2--3 is even higher than the
peak at $z \sim$ 1 suggested by optical observations,
the prediction 
is in good agreement not only with the {\sl COBE}/FIRAS CBR measurements
but also with the counts and redshift estimation for submillimeter
sources revealed by SCUBA. In our model the dusty starbursts
in primordial elliptical galaxies hardly contribute
to the cosmic SFR measured by optical observations.

Another result that is significantly different from previous studies
is the smaller ratio of MIR/FIR CBR flux compared with models based
on the empirical $L_{\rm IR}$-$T_{\rm dust}$ relation of local
infrared galaxies (e.g., Stecker 2000). It should be noted that
our model does include the warm/nonequilibrium components of 
dust emission that dominate the thermal emission from big grain dust
at the MIR region, while several previous models gave very low MIR CBR flux 
as a valley in the CBR spectrum simply because they did not take them into 
account (e.g., MacMinn \& Primack 1996; Fall, Charlot, \& Pei 1996).
The reason we get this result is the trend that 
the warm/nonequilibrium components become less significant compared 
with thermal big-grain dust emission, with increasing temperature of dust.
This is inferred from the infrared spectrum of local galaxies, and
it is also expected from physical consideration.
If this trend is taken for granted for high-$z$ galaxies, 
the MIR/FIR flux ratio
of dust emission from forming elliptical galaxies should be much
smaller than that of local infrared galaxies, since the dust
temperature of starbursting elliptical galaxies is found to be very high in our
model. Therefore these galaxies dominate in the FIR peak of CBR while
they have very small contribution in MIR CBR. This effect is 
significant enough to resolve the {\it prima facie} conflict 
(Stecker 2000) between
the upper limits on MIR CBR from TeV gamma-ray observations and
FIR CBR detections by {\sl COBE}/DIRBE and FIRAS. 

The authors have financially been supported by the JSPS Fellowship.
The authors thank Hirohisa Nagata for providing their temperature data
of {\sl IRAS} galaxies.



\begin{table}
\caption{Polynomial Fits to $\tau (\epsilon_\gamma, z_s)$}
\begin{tabular}{ccccc}
\hline \hline
$j$ &  $a_{0j}$ &  $a_{1j}$ &    $a_{2j}$ &   $a_{3j}$ \\
\hline
  0 &    1.11   &    0.85   &     $-0.28$   &   0.32     \\
  1 &    1.16   &    0.42   &      0.87   &  $-0.53$     \\
  2 &    0.04   &    0.21   &      0.35   &  $-0.23$     \\
\hline \hline
\end{tabular}
\\
Note.--- This fitting formula is accurate within 10\% at $ z_s = 0.03$
and $1 \leq \epsilon_\gamma \leq 20$, and within $\sim 50$\% 
for ranges of $0.03 \leq z_s \leq 0.3$ and
$0.1 \leq \epsilon_\gamma \leq 50$ TeV.
\label{table:TeVopt}
\end{table}

\newpage

\epsscale{1.0}

\twocolumn

\clearpage

\onecolumn
\epsscale{0.8}

\begin{figure}
\plotone{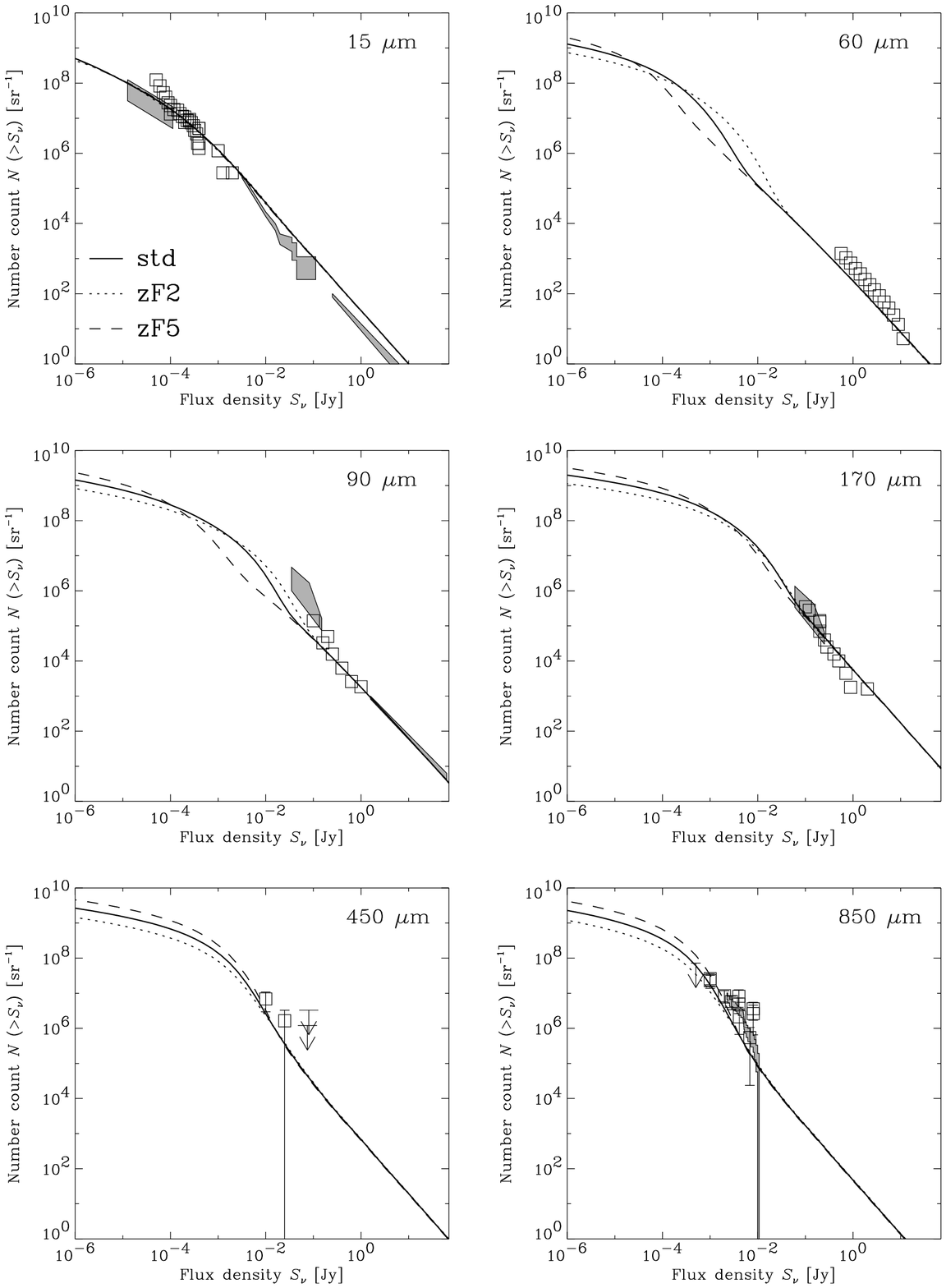}
\caption{The same as Fig. \ref{fig:nc-std}, but for different
formation redshift of galaxies: $z_F = 2$ (dotted), $z_F = 3$
(solid, the baseline model shown in Fig. \ref{fig:nc-std}), and
$z_F = 5$ (dashed).
}
\label{fig:nc-zF}
\end{figure}

\clearpage
\twocolumn
\epsscale{1.0}

\end{document}